\documentclass[narrowdisplay,onecolumn]{elsart3p}
\usepackage{graphics}
\usepackage{lineno}
\usepackage{amsmath}
\usepackage{amssymb}
\usepackage{epsfig}

\newcommand{\D}{\cdot\cdot\cdot}
\newcommand{\La}{{\Lambda}}
\newcommand{\Si}{{\Sigma}}
\newcommand{\be}{\begin{eqnarray}}
\newcommand{\ee}{\end{eqnarray}}

\begin{document}

\begin{frontmatter}
\title{Exotic bound states of two baryons in light of\\ chiral effective field theory}
\author[1]{J. Haidenbauer},
\author[1,2]{U.-G. Mei\ss ner}
\address[1]{Institute for Advanced Simulation, Institut f{\"u}r Kernphysik (Theorie) and
J\"ulich Center for Hadron Physics, Forschungszentrum J{\"u}lich, D-52425 J{\"u}lich, Germany}
\address[2]{Helmholtz-Institut f{\"u}r Strahlen- und Kernphysik (Theorie)
and Bethe Center for Theoretical Physics, Universit{\"a}t Bonn, D-53115 Bonn, Germany}

\begin{abstract}
Baryon-baryon bound states in the strangeness $S=-2$, $S=-3$, 
and $S=-4$ sectors are considered.
In particular, the dependence of the corresponding binding energies
on the quark mass (or equivalently the pion mass) is explored in the 
framework of chiral effective field theory, in order to connect 
with current lattice QCD calculations.
For a bound state in the $\Xi\Xi$ $^1S_0$ channel, predicted by our 
leading-order effective field theory interaction, binding energies are 
inferred that are roughly in line with a recent lattice QCD result 
at meson and baryon masses that correspond to those in the lattice 
simulation. 
With regard to the so-called $H$-dibaryon it is shown that the SU(3) 
breaking effects induced by the 
differences of the pertinent two-baryon thresholds ($\La\La$, $\Xi N$, 
$\Si\Si$) have a very pronounced impact on its binding energy.
A bound $H$-dibaryon as found in two lattice calculations could be 
shifted above the $\La\La$- or even above the $\Xi N$ threshold for
physical masses of the involved baryons. 
\end{abstract}
\end{frontmatter}

\section{Introduction}

The term dibaryon is used somewhat ambiguously in the literature.
It is applied for single hadrons viewed as genuine compact 
six-quarks states that are tied together
by (rather short-ranged) gluon-exchange forces between the quarks, on
one hand side, but also for loosely bound two-baryons systems 
such as the deuteron, 
that are formed by long-ranged forces between their constituents.
Possibly the most famous one of the former kind
is the $H$-dibaryon which was predicted by Jaffe in
1977 as a deeply bound state with quantum numbers of the 
$\Lambda\Lambda$ system, i.e. strangeness ${S}=-2$ and isospin $I=0$, 
and with $J^P=0^+$ \cite{Jaffe:1976yi}. 

In any case, the aforementioned deuteron (the neutron-proton bound state in 
the $^3S_1$-$^3D_1$ channel) is so far the only known and unambiguously 
established dibaryon. The interaction in the $^1S_0$ partial wave
of the neutron-proton ($np$) system is just not strong enough to 
produce a bound state and only a virtual state is created. 
Certainly, there is no shortage of new proposals of
dibaryon candidates in nucleon-nucleon scattering \cite{Adlarson} 
as well as in the strangeness sector \cite{Gal10}. 
In particular, the (approximate) SU(3) flavor symmetry of the strong 
interaction suggests that bound states could exist also in other systems 
formed by two octet baryons \cite{Miller}. 
Indeed meson-exchange models like the Nijmegen baryon-baryon ($BB$) 
interaction \cite{Stoks:1999bz}, derived under the assumption of (broken) SU(3) 
symmetry, predict bound states for the $\Xi\Xi$ but also for the 
$\Xi\Sigma$ and $\Xi\Lambda$ systems. 
A $BB$ interaction derived in a rather different way, namely within the 
framework of chiral effective field theory (EFT) \cite{Polinder06,Polinder07}
generates likewise bound states in the strangeness $S=-3$ and $S=-4$ 
sectors \cite{Hai10a}. 

With regard to the $H$-dibaryon, many experimental searches were carried out 
over the years, but so far no convincing signal was found \cite{Yoon}.
Recently, however, the $H$-dibaryon was put back on the agenda by lattice 
QCD calculations performed by the NPLQCD \cite{Beane,Beane11a} and HAL QCD 
\cite{Inoue,Inoue11a} Collaborations, where evidence for a bound state in 
the pertinent $BB$ channel was found. 
The NLPQCD Collaboration reported also evidence for a $\Xi^-\Xi^-$ bound 
state \cite{Beane11a}.
Nevertheless, one has to keep in mind that most present-day lattice QCD calculations 
are not performed at the physical masses of the involved particles. Thus, it 
is an open question how the binding energies of the calculated states evolve when 
those masses approach their physical values. Standard chiral extrapolations 
\cite{Beane11,Shanahan11} might reach their 
limits in case of dynamically generated bound states where there is a 
delicate interplay between the interaction potential (that depends on the pion mass)
and the kinetic energy (that is affected by the baryon masses). Specifically
in situations where two or more $BB$ channels can couple, as it is the case for 
the $H$-dibaryon, the effects due to the baryon masses could be sizeable.  

In this paper, we analyze various issues related to lattice QCD calculations
in the framework of chiral effective field theory  for the $BB$ 
interaction at leading order (LO) in the Weinberg counting. 
Indeed, the framework of chiral effective field theory in which our 
$BB$ interactions are derived is very well suited to shed light 
on the general characteristics of possible dibaryon bound states 
and, in particular, to study the 
quark mass\footnote{Because of the Gell-Mann-Oakes-Renner relation, the pion mass 
squared is proportional to the average light quark mass. Therefore, the notions 
``quark mass dependence'' and ``pion mass dependence'' can be used synonymously.} 
dependence of the binding energies of those states,  
in complete analogy to calculations of the quark mass dependence of the deuteron
binding energy performed in Refs.~\cite{Beane:2002vs,Beane03,Epe02,Epe02a}. 
Another important issue that can be addressed here is how this quark mass 
dependence is affected when the SU(3) breaking manifested in the masses
of the octet baryons is accounted for.

The imposed (approximate) SU(3) flavor symmetry fixes the interactions
in the $S=-3$ and $S=-4$ sectors uniquely, once the (five) low-energy 
constant (LECs) that occur at LO in chiral EFT are determined 
by a fit to the available hyperon-nucleon data.  
In particular, our LO interaction published in \cite{Polinder06}
implies the existence of several bound states in those systems. 
It will be interesting to see how the corresponding binding energies
evolve when we increase the pion mass in order to match with the
conditions of present lattice QCD calculations 
\cite{Beane,Beane11a,Inoue,Inoue11a}.

In the $S=-2$ sector with isospin zero where the $H$-dibaryon is expected 
there is one additional LEC, corresponding to the SU(3) flavor-singlet channel, 
that can not be fixed by hyperon-nucleon data. 
Since the scarce experimental information available for this sector
($\Xi^-p \to \Xi^-p$ and $\Xi^-p \to \Lambda\Lambda$ 
cross sections \cite{Ahn:2005jz}) is afflicted with large uncertainties and
does not allow to constrain its value \cite{Polinder07},
one can exploit this freedom and fine-tune the remaining LEC to produce a 
bound $H$ with a given binding energy, and then study its properties 
\cite{Haidenbauer11}.
The case of the $H$-dibaryon is also very well suited to examine
the effects from the SU(3) breaking in the baryon masses 
because, as said before, for the quantum numbers in question there are three
baryon-baryon channels that can couple, 
namely $\Lambda\Lambda$, $\Xi N$, and $\Sigma\Sigma$.
Their physical thresholds are well separated, whereas in a completely SU(3) 
symmetric world all $BB$ thresholds are degenerate. 
We will see that this has very definite dynamical consequences. 

Our manuscript is organized as follows: In Sec.~\ref{sec:2}, we recall the
basic formalism of the $BB$ interaction in the framework of chiral EFT. 
Sec.~\ref{sec:3} contains a detailed discussion of the quark 
mass dependence of binding energies in the strangeness $S=-3$ and $S=-4$ 
sectors, where our chiral EFT interaction but also the meson-exchange potential
of the Nijmegen group predict bound states in several $BB$ channels.
In Sec.~\ref{sec:4} we discuss in detail the situation for
the $H$ dibaryon. Specifically, we examine the influence of the SU(3) 
breaking through the various two-baryon thresholds and we try to 
make direct contact to the results published by the NPLQCD and 
HAL QCD Collaborations.
The paper ends with some concluding remarks. 

\section{The baryon-baryon interaction to leading order}
\label{sec:2}

For details on the derivation of the chiral $BB$ potentials for the strangeness sector 
at LO using the Weinberg power counting, we refer the reader to 
Refs.~\cite{Polinder06,Polinder07,Haidenbauer07},
see also Refs.~\cite{Savage1,Korpa,Savage2}.
Here, we just briefly summarize the basic ingredients of the chiral EFT for $BB$ interactions. 
\begin{table*}[t]
\renewcommand{\arraystretch}{1.2}
\centering
\begin{tabular}{|l|c|c|l|c|l|}
\hline
&Channel &Isospin &$C_{1S0}$ &Isospin &$C_{3S1}$\\
\hline
$S=0$&$NN\rightarrow NN$ &$1$ & $C^{27}$ &$0$ &$C^{10^*}$\\
\hline
$S=-1$&$\Lambda N \rightarrow \Lambda N$ &$\frac{1}{2}$ &$\frac{1}{10}\left(9C^{27}+C^{8_s}\right)$
&$\frac{1}{2}$ &$\frac{1}{2}\left(C^{8_a}+C^{10^*}\right)$\\
&$\Lambda N \rightarrow \Sigma N$ &$\frac{1}{2}$ &$\frac{3}{10}\left(-C^{27}+C^{8_s}\right)$
&$\frac{1}{2}$ &$\frac{1}{2}\left(-C^{8_a}+C^{10^*}\right)$\\
&$\Sigma N \rightarrow \Sigma N$ &$\frac{1}{2}$ &$\frac{1}{10}\left(C^{27}+9C^{8_s}\right)$
&$\frac{1}{2}$ &$\frac{1}{2}\left(C^{8_a}+C^{10^*}\right)$\\
&$\Sigma N \rightarrow \Sigma N$ &$\frac{3}{2}$ &$C^{27}$
&$\frac{3}{2}$ &$C^{10}$\\
\hline
$S=-2$&$\Lambda\Lambda \rightarrow \Lambda\Lambda$ &$0$ & $\frac{1}{40}\left(27C^{27}+8C^{8_s}+5C^{1}\right)$
  & & \\
&$\Lambda\Lambda \rightarrow \Xi N$ &$0$ &$\frac{-1}{40}\left(18C^{27}-8C^{8_s}-10\,C^{1}\right)$
  & & \\
&$\Lambda\Lambda \rightarrow \Sigma\Sigma$ &$0$ &$\frac{\sqrt{3}}{40}\left(-3C^{27}+8C^{8_s}-5C^{1}\right)$
  & & \\
&$\Xi N \rightarrow \Xi N$ &$0$ &$\frac{1}{40}\left(12C^{27}+8C^{8_s}+20\,C^{1}\right)$
  &$0$ &$C^{8_a}$\\
&$\Xi N \rightarrow \Sigma\Sigma$ &$0$ &$\frac{\sqrt{3}}{40}\left(2C^{27}+8C^{8_s}-10\,C^{1}\right)$
  &$1$ &$\frac{\sqrt{2}}{6}\left(C^{10}+C^{10^*}-2C^{8_a}\right)$\\
&$\Sigma\Sigma \rightarrow \Sigma\Sigma$ &$0$ &$\frac{1}{40}\left(C^{27}+24C^{8_s}+15C^{1}\right)$
 &$1$ &$\frac{1}{6}\left(C^{10}+C^{10^*}+4C^{8_a}\right)$\\
&$\Xi N \rightarrow \Xi N$ &$1$ &$\frac{1}{5}\left(2C^{27}+3C^{8_s}\right)$
  &$1$ &$\frac{1}{3}\left(C^{10}+C^{10^*}+C^{8_a}\right)$\\
&$\Xi N \rightarrow \Sigma\Lambda$ &$1$ &$\frac{\sqrt{6}}{5}\left(C^{27}-C^{8_s}\right)$
  &$1$ &$\frac{\sqrt{6}}{6}\left(C^{10}-C^{10^*}\right)$\\
&$\Sigma\Lambda \rightarrow \Sigma\Lambda$ &$1$ &$\frac{1}{5}\left(3C^{27}+2C^{8_s}\right)$
  &$1$ &$\frac{1}{2}\left(C^{10}+C^{10^*}\right)$\\
&$\Sigma\Lambda \rightarrow \Sigma\Sigma$ &   &   &$1$ &$\frac{\sqrt{3}}{6}\left(C^{10}-C^{10^*}\right)$\\
&$\Sigma\Sigma \rightarrow \Sigma\Sigma$ &$2$ &$C^{27}$
  & & \\
\hline
$S=-3$&$\Xi\Lambda \rightarrow \Xi\Lambda $ &$\frac{1}{2}$ &$\frac{1}{10}\left(9C^{27}+C^{8_s}\right)$
&$\frac{1}{2}$ &$\frac{1}{2}\left(C^{8_a}+C^{10}\right)$\\
&$\Xi \Lambda \rightarrow \Xi \Sigma $ &$\frac{1}{2}$ &$\frac{3}{10}\left(-C^{27}+C^{8_s}\right)$
&$\frac{1}{2}$ &$\frac{1}{2}\left(-C^{8_a}+C^{10}\right)$\\
&$\Xi \Sigma \rightarrow \Xi \Sigma $ &$\frac{1}{2}$ &$\frac{1}{10}\left(C^{27}+9C^{8_s}\right)$
&$\frac{1}{2}$ &$\frac{1}{2}\left(C^{8_a}+C^{10}\right)$\\
&$\Xi \Sigma \rightarrow \Xi \Sigma $ &$\frac{3}{2}$ &$C^{27}$
&$\frac{3}{2}$ &$C^{10^*}$\\
\hline
$S=-4$&$\Xi\Xi\rightarrow \Xi\Xi$ &$1$ & $C^{27}$ &$0$ &$C^{10}$\\
\hline
\end{tabular}
\caption{Various LO baryon-baryon contact potentials for the ${}^1S_0$ and ${}^3S_1$ partial
waves in the isospin basis. $C^{27}$ etc. refers to the corresponding ${\rm SU(3)_f}$
irreducible representation.}
\label{tab:1}
\end{table*}
\renewcommand{\arraystretch}{1.0}

The LO potential consists of four-baryon contact terms without derivatives and of 
one-pseudoscalar-meson exchanges. 
 The LO ${\rm SU(3)}_{\rm f}$ invariant contact terms for the octet $BB$
interactions that are Hermitian 
and invariant under Lorentz transformations follow from the Lagrangians
\begin{eqnarray}
{\mathcal L}^1 &=& C^1_i \left<\bar{B}_a\bar{B}_b\left(\Gamma_i B\right)_b\left(\Gamma_i B\right)_a\right>\ , \quad
{\mathcal L}^2 = C^2_i \left<\bar{B}_a\left(\Gamma_i B\right)_a\bar{B}_b\left(\Gamma_i B\right)_b\right>\ , \nonumber \\
{\mathcal L}^3 &=& C^3_i \left<\bar{B}_a\left(\Gamma_i B\right)_a\right>\left<\bar{B}_b\left(\Gamma_i B\right)_b\right>\  .
\label{eq:2.1}
\end{eqnarray}
Here $a, b$ denote the Dirac indices of the particles, $B$ is the irreducible octet (matrix) 
representation of ${\rm SU(3)}_{\rm f}$, and the $\Gamma_i$ are the usual elements of the 
Clifford algebra \cite{Polinder06}. As described in Ref.~\cite{Polinder06}, 
to LO the Lagrangians in Eq.~(\ref{eq:2.1}) give rise to only six independent 
low-energy constants (LECs), the $C_i^j$ in Eq.~(\ref{eq:2.1}), due to 
${\rm SU(3)}_{\rm f}$ constraints. They need to be determined by a fit to experimental data. 
It is convenient to re-express the $BB$ potentials in terms of the ${\rm SU(3)_f}$ 
irreducible representations, see e.g. Refs.~\cite{Swart,Dover}.
Then the contact interaction is given by
\begin{equation}
V=
\frac{1}{4}(1-\mbox{\boldmath $\sigma$}_1\cdot \mbox{\boldmath $\sigma$}_2) \, C_{1S0}
+ \frac{1}{4}(3+\mbox{\boldmath $\sigma$}_1 \cdot\mbox{\boldmath $\sigma$}_2) \, C_{3S1} \ ,
\label{contact}
\end{equation}
and the constraints imposed by the assumed ${\rm SU(3)}_{\rm f}$ symmetry on the interactions
in the various $BB$ channels for the $^1S_0$ and $^3S_1$ partial waves can be
readily read off from Table~\ref{tab:1}.
 
The lowest order ${\rm SU(3)}_{\rm f}$ invariant pseudoscalar-meson--baryon
interaction Lagrangian embodying the appropriate symmetries was also discussed in \cite{Polinder06}. 
The invariance under ${\rm SU(3)}_{\rm f}$ 
transformations implies specific relations between the various coupling constants, namely
\begin{equation}
\begin{array}{rlrlrl}
f_{NN\pi}  = & f, & f_{NN\eta_8}  = & \frac{1}{\sqrt{3}}(4\alpha -1)f, & f_{\Lambda NK} = & -\frac{1}{\sqrt{3}}(1+2\alpha)f, \\
f_{\Xi\Xi\pi}  = & -(1-2\alpha)f, &  f_{\Xi\Xi\eta_8}  = & -\frac{1}{\sqrt{3}}(1+2\alpha )f, & f_{\Xi\Lambda K} = & \frac{1}{\sqrt{3}}(4\alpha-1)f, \\
f_{\Lambda\Sigma\pi}  = & \frac{2}{\sqrt{3}}(1-\alpha)f, & f_{\Sigma\Sigma\eta_8}  = & \frac{2}{\sqrt{3}}(1-\alpha )f, & f_{\Sigma NK} = & (1-2\alpha)f, \\
f_{\Sigma\Sigma\pi}  = & 2\alpha f, &  f_{\Lambda\Lambda\eta_8}  = & -\frac{2}{\sqrt{3}}(1-\alpha )f, & f_{\Xi\Sigma K} = & -f.
\end{array}
\label{su3}
\end{equation}
Here $f\equiv g_A/2F_\pi$, where $g_A$ is the nucleon axial-vector strength
and $F_\pi$ is the weak pion 
decay constant.  We use the values $g_A= 1.26$ and $F_\pi = 92.4$~MeV.
For $\alpha$, the $F/(F+D)$-ratio \cite{Polinder06}, we adopt 
the SU(6) value: $\alpha=0.4$, which is consistent with recent determinations
of the axial-vector coupling constants \cite{Ratcliffe,Yamanishi}.

The spin-space part of the LO one-pseudoscalar-meson-exchange potential is similar to the 
static one-pion-exchange potential in chiral EFT for nucleon-nucleon
interactions, see e.g. \cite{Epe98} (recoil and relativistic corrections give 
higher order contributions),
\begin{eqnarray}
V^{B_1B_2\to B_1'B_2'}&=&-f_{B_1B_1'P}f_{B_2B_2'P}\frac{\left(\mbox{\boldmath $\sigma$}_1\cdot{\bf q}\right)\left(\mbox{\boldmath $\sigma$}_2\cdot{\bf q}\right)}{{\bf q}^2+M^2_P}\ ,
\label{eq:14}
\end{eqnarray}
where $M_P$ is the mass of the exchanged pseudoscalar meson. The transferred 
momentum ${\bf q}$ is defined in terms of the final and initial 
center-of-mass (c.m.) momenta of the baryons, ${\bf p}'$ and ${\bf p}$, as 
${\bf q}={\bf p}'-{\bf p}$. 
In the calculation we use the (isospin averaged) physical masses of the exchanged 
pseudoscalar mesons, i.e.
$M_\pi = 138.04\,$MeV, $M_K = 495.66\,$MeV, and $M_\eta = 548.8\,$MeV.
The explicit ${\rm SU(3)}$ breaking reflected in the mass splitting between the 
pseudoscalar mesons and, in particular, the small mass of the pion relative to
the other members of the octet leads to sizeable differences in the range of
the interactions in the different channels and, thus, induces an essential dynamical 
breaking of ${\rm SU(3)}$ symmetry in the $BB$ interactions.
The $\eta$ meson was identified with the octet $\eta$ ($\eta_8$) and its physical 
mass was used. Note that for getting the actual potential for a specific channel
one still has to multiply the expression in Eq.~(\ref{eq:14}) with the pertinent
isospin coefficient (as given, e.g., in Ref.~\cite{Polinder06}).

The reaction amplitudes are obtained from the solution of a coupled-channel 
Lippmann-Schwinger (LS) equation for the interaction potentials: 
\begin{eqnarray}
&&T_{\rho''\rho'}^{\nu''\nu',J}(p'',p';\sqrt{s})=V_{\rho''\rho'}^{\nu''\nu',J}(p'',p')+
\sum_{\rho,\nu}\int_0^\infty \frac{dpp^2}{(2\pi)^3} \, V_{\rho''\rho}^{\nu''\nu,J}(p'',p)
\frac{2\mu_{\nu}}{q_{\nu}^2-p^2+i\eta}T_{\rho\rho'}^{\nu\nu',J}(p,p';\sqrt{s})\ .
\label{LS} 
\end{eqnarray}
The label $\nu$ indicates the particle channels and the label $\rho$ the partial wave. 
$\mu_\nu$ is the pertinent reduced mass. The on-shell momentum in the intermediate state, 
$q_{\nu}$, is defined by $\sqrt{s}=\sqrt{m^2_{B_{1,\nu}}+q_{\nu}^2}+\sqrt{m^2_{B_{2,\nu}}+q_{\nu}^2}$. 
Relativistic kinematics is used for relating the laboratory energy $T_{{\rm lab}}$ of the hyperons 
to the c.m. momentum.

In \cite{Polinder06,Polinder07} 
the LS equation was solved in the particle basis, in order to incorporate the correct physical
thresholds. Since here we are primarily interested in bound states we work in the isospin
basis. Furthermore, we ignore the Coulomb interaction (as it is also done in
the pertinent lattice QCD calculations). 
We use the following (isospin averaged) baryon masses: $m_N=939.6$ MeV, 
$m_\La=1115.6$ MeV, $m_\Si=1192.5$ MeV, and $m_\Xi=1318.1$ MeV.
In the $S=-4$ and $S=-3$ sectors either single channel ($\Xi\Xi$) or 
coupled-channel ($\Xi\Lambda - \Xi\Sigma$) equations have to be solved. For $S=-2$ and,
in particular, for the $H$-dibaryon there are three coupled channels, namely $\La\La$, $\Xi N$ 
and $\Si\Si$. The potentials in the LS 
equation are cut off with a regulator function, $\exp\left[-\left(p'^4+p^4\right)/\Lambda^4\right]$, 
in order to remove high-energy components of the baryon and pseudoscalar meson fields \cite{Epe05}.
We consider cut-off values in the range from 550 to 700 MeV, similar to what was used for  
chiral $NN$ potentials \cite{Epe05}.

The imposed ${\rm SU(3)}$ flavor symmetry implies that only five of the six LECs 
contribute to the $YN$ interaction, namely $C^{27}$, $C^{10}$, $C^{10^*}$, $C^{8_s}$, 
and $C^{8_a}$, cf. Table~\ref{tab:1}. 
These five contact terms were determined in 
\cite{Polinder06} by a fit to the $YN$ scattering data. Since the $NN$ data
cannot be described with a LO EFT (except very close to the threshold), 
${\rm SU(3)}$ constraints from the $NN$ interaction 
were not implemented explicitly. As shown in Ref.~\cite{Polinder06},  a good
description  of the 35 low-energy $YN$ scattering can be 
obtained for cutoff values $\Lambda=550,...,700$ MeV and for natural values of the LECs. 
The sixth LEC ($C^{1}$) is only present in the $S=-2$ channels with isospin zero,
cf. Table~\ref{tab:1}. There is scarce experimental information on these 
channels that could be used to fix this LEC, but it turned out that the quality 
of the existing data does not really allow to constrain its value reliably 
\cite{Polinder07}. Even with the value of the sixth LEC chosen so that 
$C^{\La\La\to \La\La}_{1S0} = 0$, agreement with those data can be achieved. In this case 
a scattering length of $a_{^1S_0}^{\Lambda\Lambda} = -1.52$~fm \cite{Polinder07} 
is obtained. 
Analyses of the measured binding energy of the double-strange hypernucleus
${}^{\;\;\;6}_{\Lambda\Lambda}{\rm He}$ \cite{Takahashi:2001nm} suggest that
the $\La\La$ scattering length could be in the range of 
$-1.3$ to $-0.7$ fm \cite{Gal,Rijken,Fujiwara}.
A first determination of the scattering length utilizing data on the
$\Lambda\Lambda$ invariant mass from the reaction
$^{12}C(K^-,K^+\Lambda\Lambda X)$ led to the result
$a^{\Lambda\Lambda}=-1.2\pm 0.6$ fm \cite{Ashot}. 
 
\section{Quark mass dependence of baryon-baryon binding energies}
\label{sec:3}

As discussed in Ref.~\cite{Hai10a}, our LO chiral EFT interaction 
predicts several bound states for the strangeness $S=-3$ and $S=-4$
sectors.
Let us start with the ${}^1S_0$ partial wave in the $\Xi^0\Lambda$ 
channel. For the smallest cut-off ($\Lambda = 550$ MeV) only a virtual 
state is found in this partial wave which, however, 
eventually transforms into a real bound state when the cut-off is increased 
within the considered range. For the largest cut-off (700 MeV) a binding 
energy of $-0.43\,$MeV is predicted. 
The results for the $\Xi^0\Lambda$ channel of other potentials that provide
detailed results for the $S=-3$ and $S=-4$ sectors \cite{Stoks:1999bz,Fujiwara}
suggest also an overall attractive interaction in the ${}^1S_0$ partial 
wave though only a very moderate one which does not support a bound state. 

The $S$-waves in the $\Xi\Sigma$ $I=3/2$ channel belong to the same
($10^*$ and $27$, respectively, cf. Table~\ref{tab:1}) 
irreducible representations where in the $NN$ case bound states 
(${}^3S_1$-$^3D_1$) or virtual states (${}^1S_0$) exist.
Therefore, one expects that such states can also occur for $\Xi\Sigma$.
Indeed, here bound states are present for both partial waves in the Nijmegen 
model, cf. the discussion in Sect.~III.B in Ref.~\cite{Stoks:1999bz}. 
The chiral EFT interaction has a bound state too for ${}^1S_0$, 
for all cut-off values \cite{Hai10a}. The binding energies lie in 
the range of $-2.23\,$MeV ($\Lambda = 550\,$MeV) to $-6.15\,$MeV (700 MeV).
But in the ${}^3S_1$-$^3D_1$ partial wave the attraction is obviously not 
strong enough to form a bound state. 
The ${}^1S_0$ state of the $\Xi\Xi$ channel belongs also to the $27$plet
irreducible representation and also here the Nijmegen as well as the
chiral EFT interactions produce bound states. In our case the binding energy
lies in the range of $-2.56\,$MeV ($\Lambda = 550\,$MeV) to $-7.28\,$MeV
(700 MeV). 

\begin{figure}[t!]
\centering
\includegraphics[width=0.305\textwidth,keepaspectratio,angle=-90]{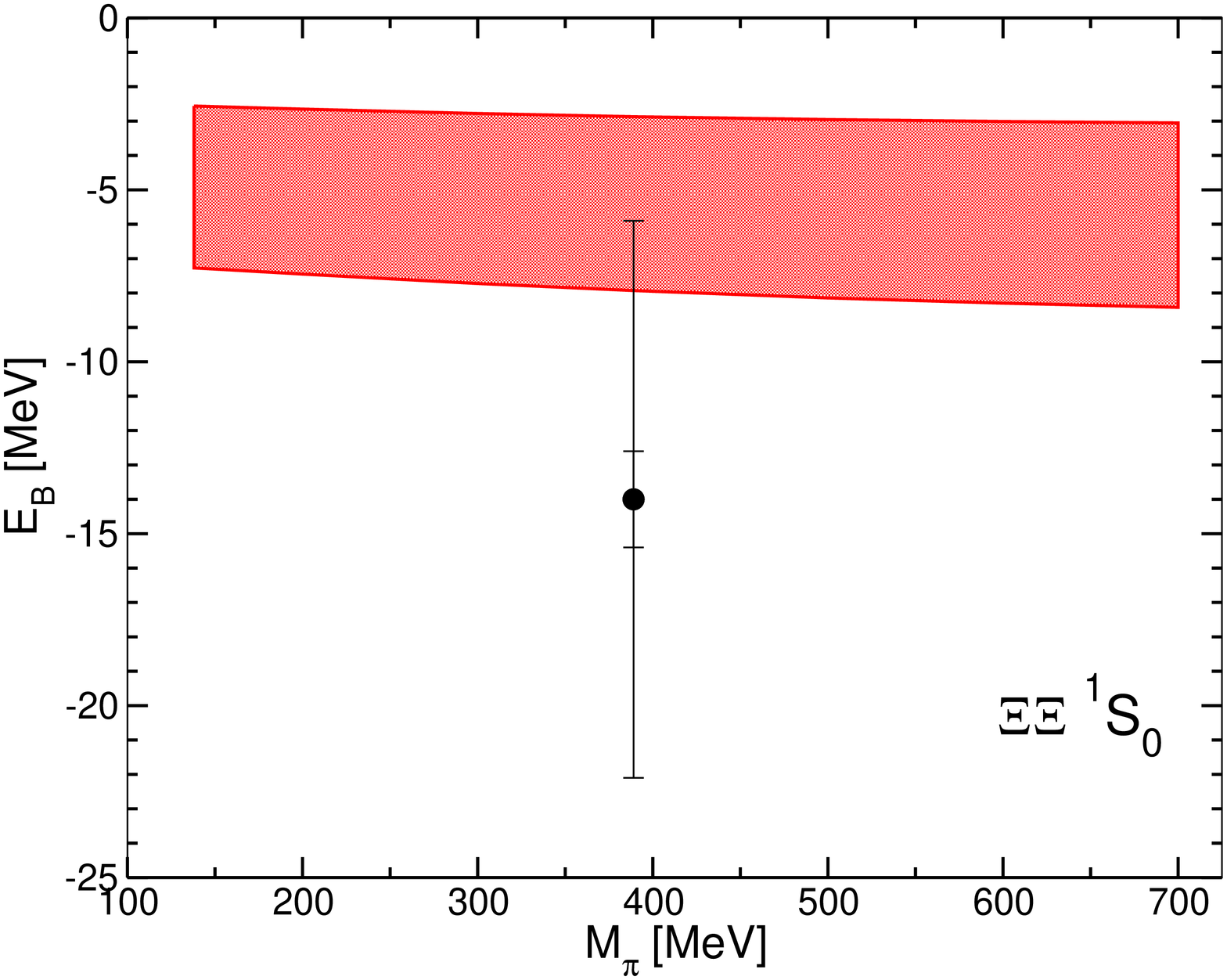}
\includegraphics[width=0.305\textwidth,keepaspectratio,angle=-90]{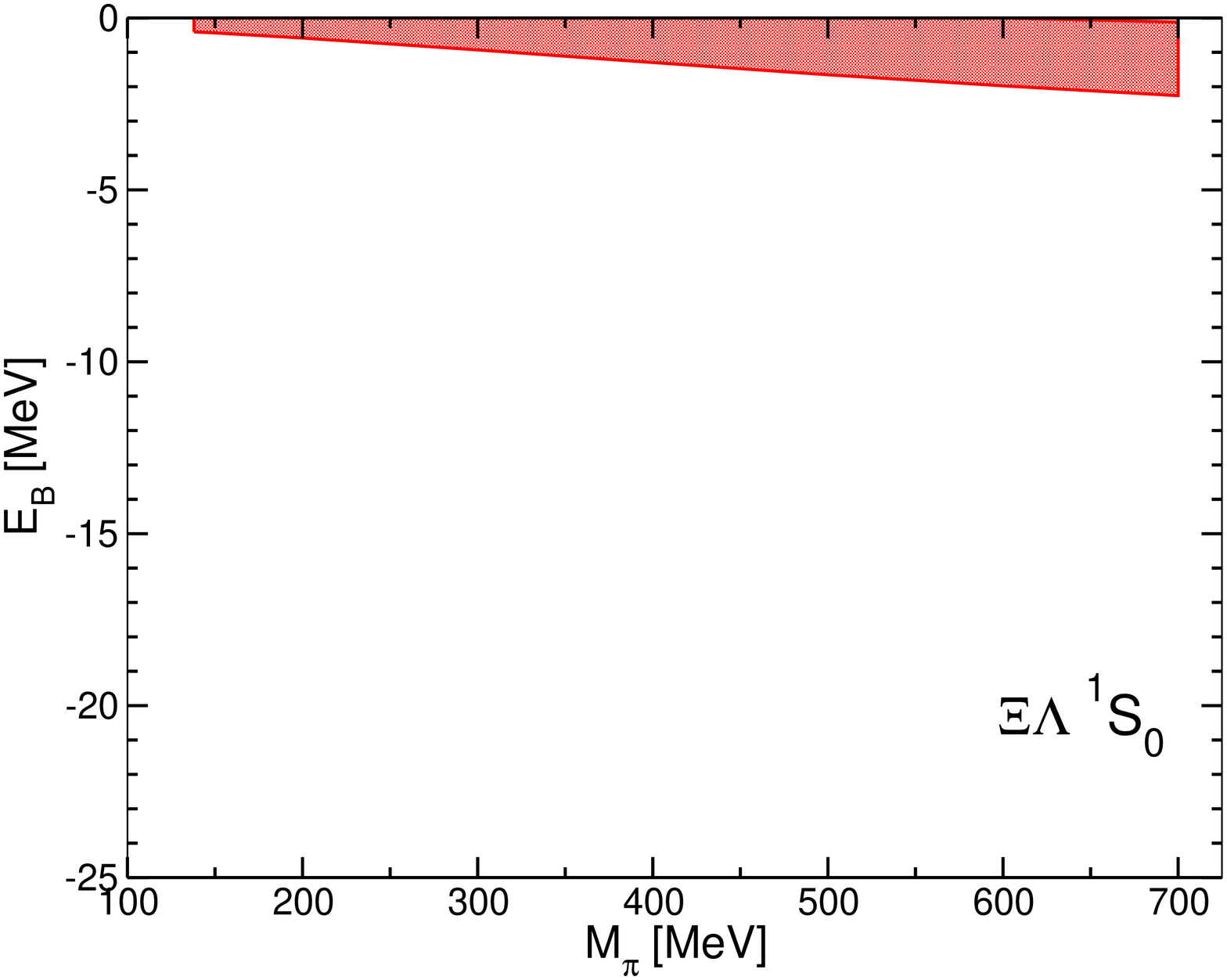}
\includegraphics[width=0.305\textwidth,keepaspectratio,angle=-90]{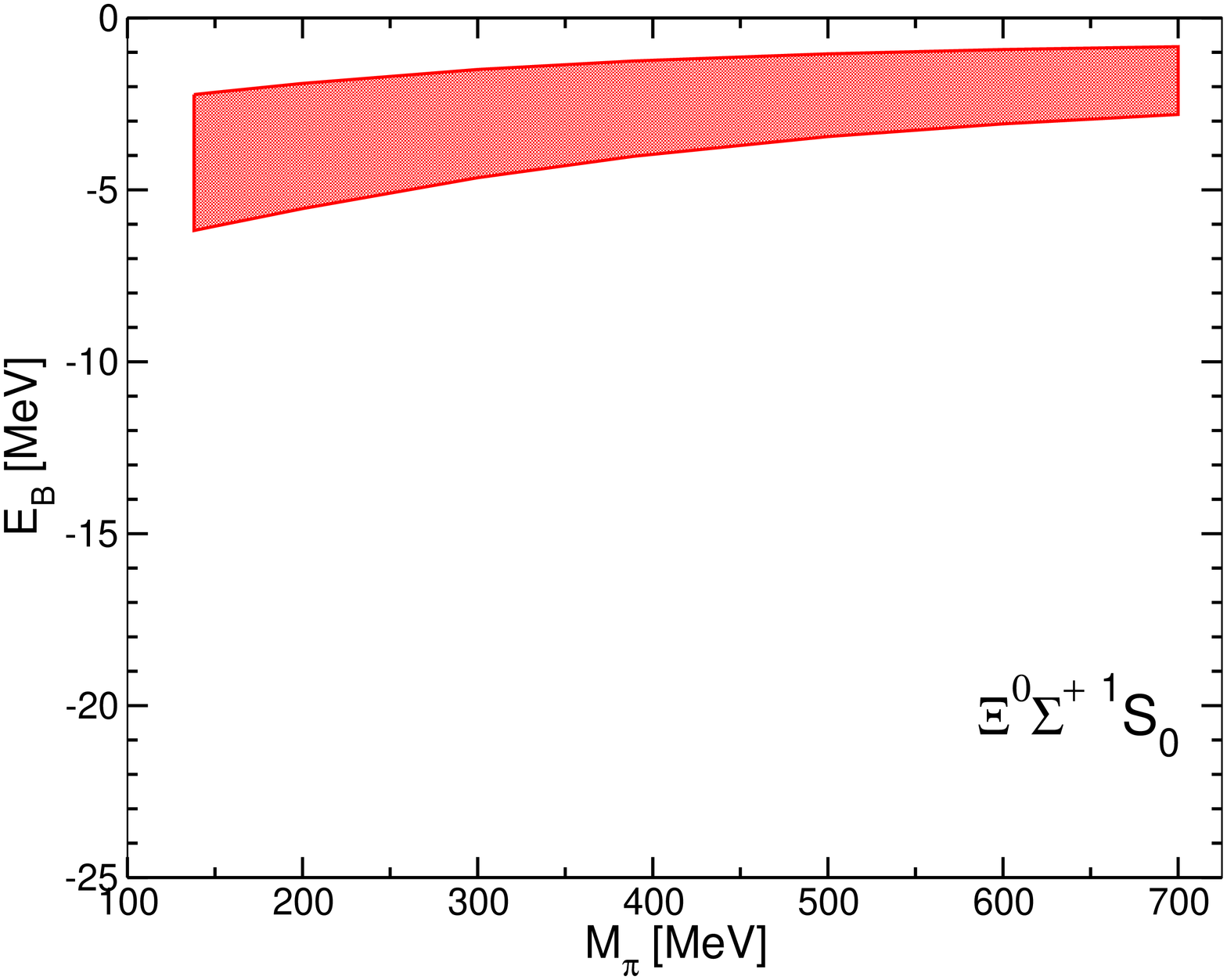}
\caption{Dependence of the binding energies in different $S=-3$ and $S=-4$ $^1S_0$ partial waves 
on the pion mass.
}
\label{fig:BE1}
\end{figure}

Since the ${}^1S_0$ partial waves in the $\Sigma N$ ($I=3/2$) and 
$\Sigma\Sigma$ ($I=2$) channels belong likewise to the $27$plet, cf. 
Table~\ref{tab:1}, one could expect bound states in those states, too.
However, our chiral EFT interaction is only moderately attractive in the 
former case, as reflected in the corresponding scattering lengths
which range from $-2.24$ to $-2.36\,$fm, cf. \cite{Polinder06}. There 
is also no bound state for $\Sigma^+\Sigma^+$, though the predicted 
scattering lengths are here between $-6.23$ to $-9.42\,$fm, which is an 
indication that there is a virtual state not too far from the physical 
region. 
Note that in our calculation 
$\Sigma \Sigma \,(I=2) \equiv \Sigma^+\Sigma^+ \equiv \Sigma^-\Sigma^-$
and, therefore, we use those designations synonymously here.
The Nijmegen NSC97 potential, on the other hand, clearly produces
a bound state in the $\Sigma^+\Sigma^+$ state, as signalled by the
large and positive scattering lengths \cite{Stoks:1999bz}. 
Interestingly, a $BB$ interaction derived within the constituent quark-model 
(fss2) \cite{Fujiwara}, yields also a scattering length that is very large 
and negative so that there should be a virtual state practically at the 
$\Sigma^+\Sigma^+$ threshold. On the other hand, for all $S=-3$ and $S=-4$ 
partial waves no bound states are predicted by this interaction model
based on quark-gluon dynamics \cite{Fujiwara}. 

Let us now consider variations of the masses of the involved particles. 
First we study the dependence of the binding energies on the pion 
mass $M_\pi$ and keep the baryon masses at their physical values. 
We will examine the specific situation for the concrete (meson and 
baryon) masses that correspond to the calculation reported by the NPLQCD 
Collaboration below. 

\begin{figure}[t!]
\centering
\includegraphics[width=0.305\textwidth,keepaspectratio,angle=-90]{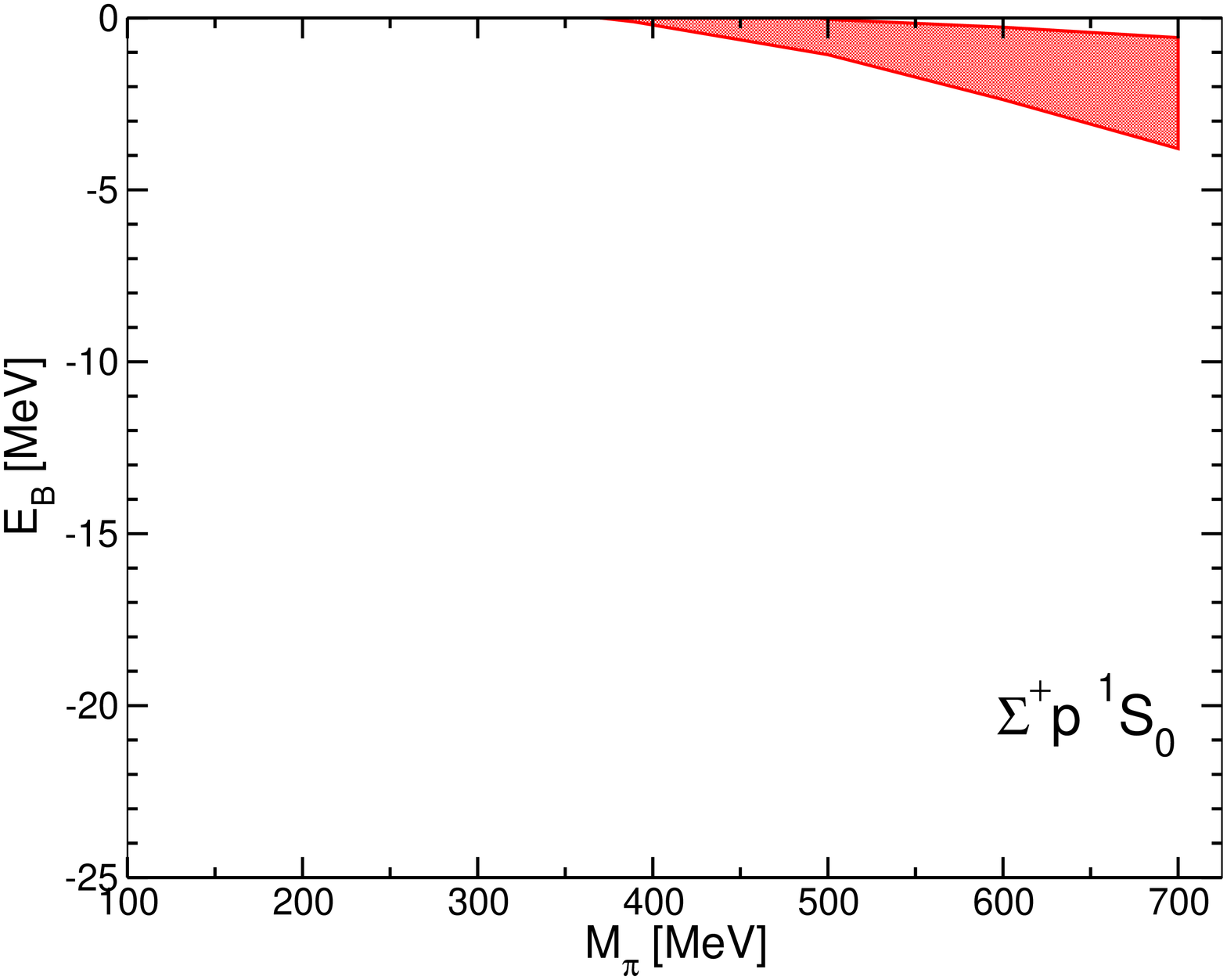}
\includegraphics[width=0.305\textwidth,keepaspectratio,angle=-90]{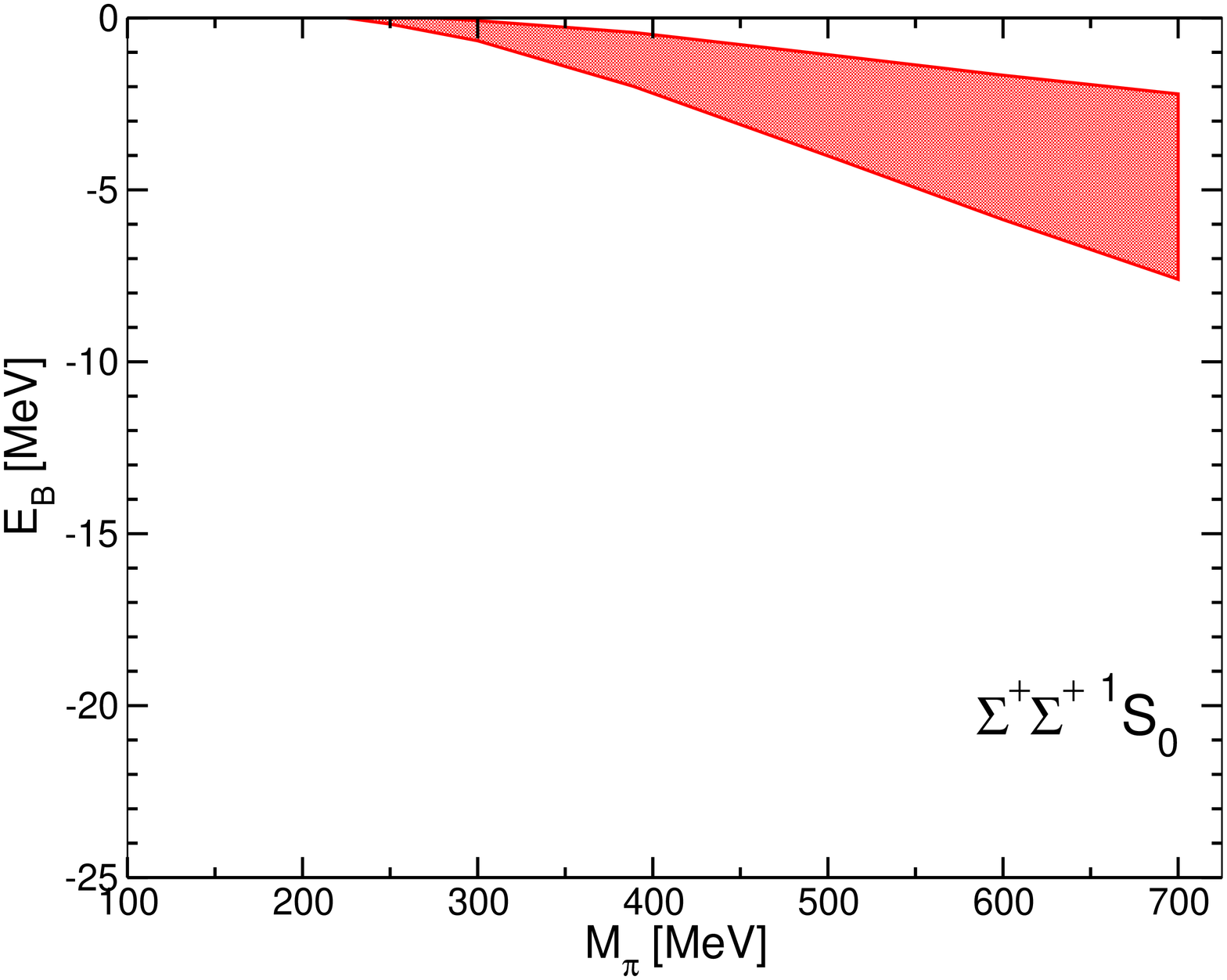}
\caption{Dependence of the binding energies in the 
$\Sigma^+ p$ (left) and $\Sigma^+ \Sigma^+$ (right) $^1S_0$ partial waves 
on the pion mass.
}
\label{fig:BE2}
\end{figure}

Our results are displayed in Figs.~\ref{fig:BE1} and \ref{fig:BE2}. 
Obviously for the $S=-3$ and $S=-4$ systems (Fig.~\ref{fig:BE1})
there is a fairly weak dependence of the predicted binding energies 
on the pion mass. In particular, the variation from the physical point
to masses around 400 MeV, corresponding to the present status of
the NPLQCD computations, are relatively small compared to the differences
due to the cut-off variations. 
Note that for $\Xi\Sigma$ 
the binding energy decreases with increasing pion mass, in contrast to
what happens in the other channels. 

Results for the $S=-1$ and $S=-2$ systems are presented in Fig.~\ref{fig:BE2}.
Contrary to the $\Lambda\Lambda$ system, which we discuss in the 
next section, the $\Sigma\Sigma$ interaction in the $I=2$ channel is
completely fixed by the five LECs that could be determined by
a fit to the $YN$ data, cf. Table~\ref{tab:1}. Thus predictions can be
made for this system too. Interestingly, while being unbound at the
physical point, a bound $\Sigma^+\Sigma^+$ state is created when the
pion mass is increased. For $M_\pi\approx$ 400 MeV the predicted 
binding energy is around 2-3 MeV. Note that corresponding investigations
within lattice QCD remained inconclusive in this case \cite{Beane11a}. 
A bound state appears too in the $\Sigma^+ p$ system, however, only for
pion masses above 400 MeV. 
Note that $\Sigma N \, (I=3/2) \equiv \Sigma^+ p \equiv \Sigma^- n$.
For both systems we observe a somewhat stronger
dependence of the binding energies on the pion mass as compared to 
$S=-3$ and $S=-4$. 

The dependence on the pion mass can be easily understood, on a 
qualitative level, by considering the contributions from
pseudoscalar-meson exchange to the interaction in the various 
baryon-baryon systems. 
Though in a fully SU(3) symmetric world 
\begin{equation} 
V^{NN\to NN}_{(I=1)} =  V^{\Sigma N\to \Sigma N}_{(I=3/2)} = 
V^{\Sigma\Sigma \to \Sigma\Sigma}_{(I=2)} = 
V^{\Xi\Sigma\to \Xi\Sigma}_{(I=3/2)} = V^{\Xi\Xi\to \Xi\Xi}_{(I=1)} = V^{27} 
\end{equation} 
in the $^1S_0$ partial wave, 
one has to keep in mind that the indivudal contributions of the pseudoscalar 
mesons differ for different channels. Their relative strengths in the various
channels follows from the product of the relevant coupling constants, fixed 
by the assumed SU(3) symmetry, which are tabulated in Eq.~(\ref{su3}), and 
a corresponding isospin factor: 

\renewcommand{\arraystretch}{1.2}
\begin{equation} 
\begin{array}{llll}
NN \to NN: \ \ & V_\pi \propto \phantom{(20/25) \times} f^2, & V_\eta \propto {(3/25)} \times f^2 & \\
\Sigma N \to \Sigma N: & V_\pi \propto {(20/25)} \times f^2, & V_\eta \propto {(6/25)} \times f^2, 
& V_K \propto {(2/25)} \times f^2 \\
\Sigma\Sigma \to \Sigma\Sigma: & V_\pi \propto {(16/25)} \times f^2, 
& V_\eta \propto {(12/25)} \times f^2 & \\
\Xi\Sigma \to \Xi\Sigma: & V_\pi \propto {(-4/25)} \times f^2, \ \ & 
V_\eta \propto {(-18/25)} \times f^2, \ \ & V_K \propto 2 \times f^2 \\
\Xi\Xi \to \Xi\Xi: & V_\pi \propto {(1/25)} \times f^2, & V_\eta \propto {(27/25)} \times f^2 & \\
\end{array}
\label{coupling} 
\end{equation}
\renewcommand{\arraystretch}{1.0}

Let us compare, for example, $NN$ and $\Xi\Xi$. Obviously in the $NN$ case 
the pion-exchange contribution dominates while
for $\Xi\Xi$ practically the whole contribution from pseudoscalar-meson
exchange is due to the $\eta$ meson. Consequently, variations of the 
pion mass (or the SU(3) breaking manifested by the small pion mass) 
are much less important for the $\Xi\Xi$ system than for 
the $NN$ interaction, cf. Refs.~\cite{Beane:2002vs,Beane03,Epe02,Epe02a} 
for a discussion of the effects in the latter system. 
Since the small value $f_{\Xi\Xi\pi}  = -0.2 \times f$ enters also into the
$\Xi\Sigma$ interaction a similarly weak dependence is seen there. Note that 
$V^{\Xi\Sigma\to \Xi\Sigma }_\pi = - 4 \times V^{\Xi\Xi \to \Xi\Xi}_\pi$ for
the isospin channels shown in Fig.~\ref{fig:BE1}, which explains the opposite
trend in the dependence of the binding energy on the pion mass. 
In case of $\Xi\Lambda$, pion-exchange contributes only via coupled-channel
effects so that one expects a weak pion-mass dependence anyway. 
In the channels $\Sigma N$ and $\Sigma\Sigma$ where the strength of 
pion exchange is less reduces as compared to $NN$ 
($V^{\Sigma N \to \Sigma N}_\pi = 4/5 \times V^{NN\to NN}_\pi$,
$V^{\Sigma \Sigma \to \Sigma \Sigma}_\pi = 16/25 \times V^{NN\to NN}_\pi$)
we observe a sizeable pion mass dependence of the binding energies, cf. 
Fig.~\ref{fig:BE2}. 
Similar comments also apply for the pion mass dependence of the
baryon octet states with increasing strangeness \cite{Frink:2005ru}.
Note that the relations in Eq.~(\ref{coupling}) follow for the SU(6)
value $\alpha = 0.4$, but they change only marginally for values of 
$\alpha \approx 0.36 - 0.37$, as determined recently in analyses of 
hyperon semi-leptonic decay data \cite{Ratcliffe,Yamanishi}. 

In order to connect as closely as possible to the results published by the
NPLQCD Collaboration \cite{Beane11b} we performed also calculations with
meson and baryon masses corresponding precisely to those in the lattice
QCD simulation. Specifically, we use $M_\pi = 389\,$MeV, 
$M_K = 544\,$MeV, and $M_\eta = 587\,$MeV, and the
baryon masses $m_N=1151.3\,$MeV, $m_\La=1241.9\,$MeV, $m_\Si=1280.3\,$MeV,
and $m_\Xi=1349.6\,$MeV, all taken from Ref.~\cite{Beane11b}. 
Corresponding results are summarized in Table \ref{BELQCD} where we also 
include the binding energies at the physical point and those where only the
pion mass was set to $M_\pi = 389\,$MeV. All values are given with two digits
behind the comma in order to facilitate an easy comparison of the relative
size of the various effects. The absolute uncertainty of our leading-order 
calculation is, of course, best reflected in the cut-off dependence of the
results represented by the shaded bands in Figs.~\ref{fig:BE1} and 
\ref{fig:BE2} and by the pertinent values in Table \ref{BELQCD}. 

The results in Table \ref{BELQCD} make clear that there are sizeable effects
from the baryon masses (and of the heavy pseudoscalar mesons $K$ and $\eta$,
too) on the binding energies. Specifically in the $S=-3$ and $S=-4$ sectors 
those are more significant than the variations in the pion mass that we 
considered, which is not surprising if one recalls the discussion above. 
Clearly, one has to acknowledge that the systematic uncertainty in 
the current lattice QCD calculations is still significantly larger 
than those mass effects \cite{Beane11a}. 
Despite of this, it is remarkable that the $\Xi^-\Xi^-$ binding 
energy published in \cite{Beane11a}, $E_B = (-14.0 \pm 1.4 \pm 6.7)\,$MeV, 
is rather well in line with the corresponding predictions based on LO chiral 
EFT. Considering the (rather modest) mass dependence found in our calculation 
we would expect that this state is still bound at the physical point,
namely by roughly 10 MeV if one takes the central value from \cite{Beane11a}
as guideline. 
Future lattice QCD calculations with improved statistics will certainly resolve 
this exciting issue, once the systematic uncertainties can be reduced. 

With regard to the other states listed in Table~\ref{BELQCD} only the one
in the $\Xi^-\Sigma^-$ $^1S_0$ partial wave is likely to survive for 
physical masses. All other states are fairly loosely bound already for masses 
corresponding to the NPLQCD calculation and disappear when we go to the
physical point. 

\begin{table}[h]
\renewcommand{\arraystretch}{1.5}
\centering
\begin{tabular}{|c|c|c|c|c|}
\hline
& \multicolumn{3}{|c|} {$\chi$EFT} & NPLQCD \cite{Beane11a} \\
\hline
                  & physical masses & $M_\pi = 389$ MeV & NPLQCD masses & \\
${\Lambda}$ [MeV] & 550 {$\D$} 700 & 550 {$\D$} 700 & 550 {$\D$} 700 & \\
\hline
\ $\Xi^-\Xi^- $         \ & \ $-2.56$ {$\D$} $-7.27$ \ & \ $-2.87$ {$\D$} $-7.93$ \ & \ $-3.92$ {$\D$} $-10.41$ \ &
 \ -14.0 $\pm$ 1.4 $\pm$ 6.7 \ \\
\ $\Xi^-\Lambda $       \ & \ $ \phantom{00}0 $ {$\D$} $-0.40$ \ & \ $ \phantom{00}0 $ {$\D$} $-1.26$ \ 
& \ $-1.05$ {$\D$} $-4.86$ \ & \\
\ $\Xi^-\Sigma^- $      \ & \ $-2.23$ {$\D$} $-6.18$ \ & \ $-1.25$ {$\D$} $-4.02$ \ & \ $-2.89$ {$\D$} $-7.93$ \
& \\
\ $\Sigma^-\Sigma^- $ \ & \ $ -  $ \ & \ $-0.42$ {$\D$} $-1.99$ \ & \ $-1.23$ {$\D$} $-3.93$ \
& inconclusive \\
\ $\Sigma^- n $       \ & \ $ -  $ \ & \ $ \phantom{00}0 \ ${$\D$} $-0.11$ \ & \ $-0.30$ {$\D$} $-1.56$ \
& \\
\hline
\end{tabular}
\caption{Binding energies in MeV of various $BB$ bound states in the $^1S_0$ partial wave 
as obtained from the EFT potential for physical masses
(second column), for a pion mass of $M_\pi = 389$ MeV (third column), 
and using meson and baryon masses that correspond to the lattice QCD calculation 
of \cite{Beane11a} (fourth column). The last column are results of the lattice
QCD calculation taken from Ref.~\cite{Beane11a}. 
}
\label{BELQCD}
\end{table}

In this context let us emphasize that, of course, it would be also interesting to 
confirm or exclude the existence of those bound states experimentally. 
The possibility to find signals for strange di-baryon states in heavy-ion collisions 
was discussed in Refs.~\cite{Schaffner1999,Steinheimer}. 
Also the new facilities J-PARC (Tokai, Japan) and FAIR (Darmstadt, Germany) 
could allow one to obtain empirical constraints on the baryon-baryon interaction 
in the $S=-3$ and $-4$ sector. Information could come from formation experiments 
of corresponding hypernuclei or from proton-proton and antiproton-proton collisions 
at such high energies that pairs of baryons with strangeness $S=-3$ or $S=-4$ can 
be produced. 

\section{The $H$-dibaryon}
\label{sec:4}

\subsection{General considerations}
As already said, in the $S=-2$ sector with isospin zero where the $H$-dibaryon 
is expected there is one additional contact term (${C^1}$, cf. Table~\ref{tab:1}),
corresponding to the SU(3) flavor-singlet channel, that is not fixed by 
hyperon-nucleon data and, therefore, no immediate predictions can be made. 
In principle, this LEC could be determined from experimental information available 
for this sector, but the scarce data 
($\Xi^-p \to \Xi^-p$ and $\Xi^-p \to \Lambda\Lambda$ 
cross sections \cite{Ahn:2005jz}) are afflicted with large uncertainties 
and do not allow to constrain its value \cite{Polinder07}.
Thus, in practice one can exploit this freedom and fine-tune the remaining 
LEC to produce a bound $H$ with a given binding energy, and then study its 
properties \cite{Haidenbauer11}. Indeed, it turned out that a near-threshold 
bound state can be easily produced for ${C^1}$ values that are of natural size.

In the following we assume that the $H$-dibaryon is a (loosely) bound $BB$ state
\cite{Haidenbauer11}, just like the bound states discussed in the previous section.
We do not 
consider the case where the $H$-dibaryon is a genuine 6-quark state as originally 
suggested by Jaffe \cite{Jaffe:1976yi}. In fact, we cannot say anything about 
the latter situation within our framework.
We also assume that the binding energy, $E_H$, is similar to 
that of the deuteron ($D$) because this allows us to compare the properties
of the generated $H$-dibaryon directly with the familiar deuteron case. 
Specifically, we fix the value of the flavor-singlet LEC ${C^1}$ in such a way that
the binding momentum is $\gamma_{H} = \gamma_{D} = 0.23161\,$fm
($E = - \gamma^2 / m_{B}$, where $m_{B}$ is either $m_N$ or $m_\Lambda$),
in view of the well-known relation between the binding energy and the effective
range parameters \cite{Schwinger47,Bethe} 
\begin{equation}
\frac{1}{a} \simeq {\gamma} - \frac{1}{2}{r} {\gamma}^2 .
\nonumber
\end{equation}
This relation is very well fulfilled for the deuteron and the corresponding
neutron-proton $^3S_1$ scattering length ($a=5.43\,$fm) and effective range
($r=1.76\,$fm). One would naively expect that the same should happen for the
$H$-dibaryon. However, it turns out that the corresponding results for
$\La\La$ in the $^1S_0$ partial wave are quite different, namely
$a=3.00\,$fm and $r=-4.95\,$fm. Specifically, the effective range
is much larger and, moreover, negative. Clearly, the
properties of the $H$-dibaryon are not comparable to those of the
deuteron, despite the fact that both bound states are close to the
elastic threshold.
Indeed, if one recalls the expressions for the relevant potentials
as given in Table \ref{tab:1},
\begin{equation}
V^{\La\La \rightarrow \La\La} = \frac{1}{40}\left(27C^{27}+8C^{8_s}+{ 5}{
    C^{1}}\right) \, , \ \ \ 
V^{\Xi N \rightarrow \Xi N} = \frac{1}{40}\left(12C^{27}+8C^{8_s}+{ 20}{
    C^{1}}\right) \, ,
\nonumber 
\end{equation}
one can see that the attraction provided by the SU(3) flavor-singlet state (i.e. $C^{1}$)
contributes with a much larger weight to the $\Xi N$ channel than to $\La\La$.
This indicates that the presumed $H$-dibaryon could be predominantly a
$\Xi N$ bound state. We have confirmed this conjecture by evaluating explicitly
the phase shifts in the $\La\La$ and $\Xi N$ channels, cf. the discussion in the
next section. Indeed, one finds that the phase shift for the $\Xi N$ channel
is rather similar to the $NN$ $^3S_1$ case. Specifically, the $\Xi N$ ($^1S_0$) phase shift 
$\delta(q_{\Xi N})$ fulfills $\delta(0)-\delta(\infty) = 180^\circ$ in 
agreement with the Levinson theorem. 
The $\La\La$ ($^1S_0$) phase behaves rather differently and 
satisfies $\delta(q_{\La\La}=0) - \delta(\infty) = 0$.
Note that there have been earlier discussions on this issue in
the context of $S=-2$ baryon-baryon interactions derived within the
quark model \cite{Oka,Nakamoto}. 

The results above were obtained with the LECs $C^{27}$ and $C^{8_s}$
fixed from the $YN$ data for the cutoff value $\Lambda = 550$~MeV \cite{Polinder06}.
We considered also the other variants corresponding to cutoff masses 
of 600, 650, and 700 MeV in the LS equation~(\ref{LS}), as in 
Ref.~\cite{Polinder07}. But since the contact term $C^{1}$ has to be 
determined anew in each case it turned out that the results are rather 
similar for all cutoffs once $C^{1}$ is fixed in such a way that 
the same binding energy for the $H$ dibaryon is produced. Thus,
we will present only results for the $\Lambda = 550$~MeV case. 
We denote the $YY$ interaction with a loosely bound $H$ dibaryon by 
YY-D in the following, and use the notation YY(550) for the 
original interaction from Ref.~\cite{Polinder07}.

\begin{figure}[t!]
\centering
\includegraphics[width=0.305\textwidth,keepaspectratio,angle=-90]{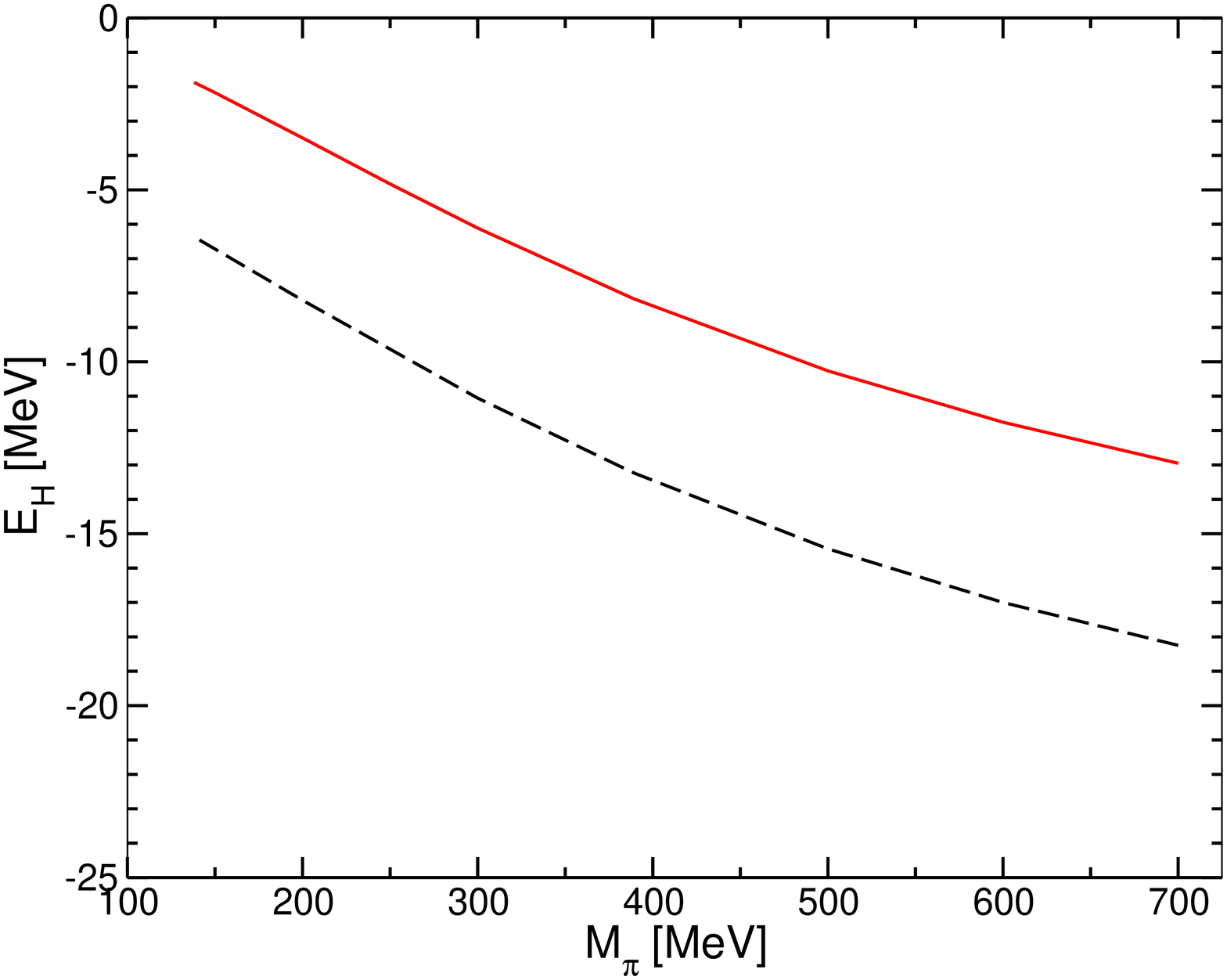}
\includegraphics[width=0.305\textwidth,keepaspectratio,angle=-90]{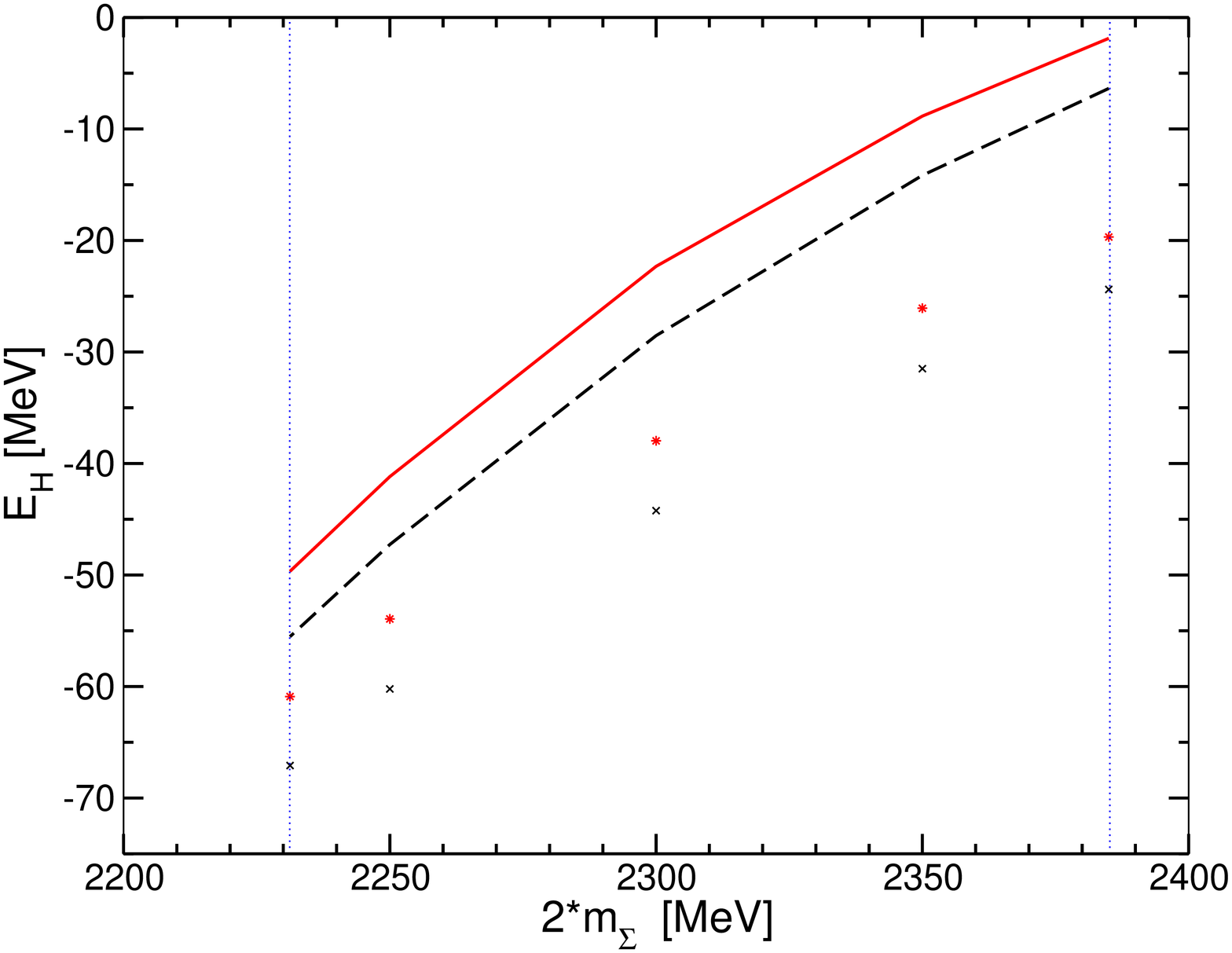}
\caption{Dependence of the binding energy of the $H$-dibaryon on the pion mass 
$M_\pi$ (left) and on the $\Sigma$ mass $m_\Sigma$ (right).
The solid curve correspond to the case where $C^1$ is fixed such that $E_H = -1.87\,$MeV 
for physical masses while for the dashed curve $C^1$ is fixed to yield $E_H = -13.2\,$MeV for 
$M_\pi = 389$ MeV. 
The asterisks and crosses represent results where, besides the 
variation of $m_\Sigma$, $m_\Xi+m_N = 2 m_\La$ is assumed so that the $\Xi N$ threshold 
coincides with that of the $\La\La$ channel. 
The vertical (dotted) lines indicate the physical $\La\La$ and $\Si\Si$ thresholds. 
}
\label{fig:mpi}
\end{figure}

Let us now consider variations of the masses of the involved particles. 
The dependence of the $H$ binding energy on the pion mass $M_\pi$ is 
displayed in Fig.~\ref{fig:mpi} (left). For the YY-D potential considered 
above, enlarging the pion mass to around 400 MeV 
(i.e. to values in an order that corresponds to the NPLQCD calculation \cite{Beane}) 
increases the
binding energy to around 8 MeV and a further change of $M_\pi$ to 700 MeV
(corresponding roughly to the HAL QCD calculation \cite{Inoue}) yields then 13 MeV, 
cf. the solid line. 
Readjusting $C^1$ so that we predict a $H$ binding energy of 13.2 MeV for
$M_\pi=389$ MeV, corresponding to the latest result published by 
NPLQCD~\cite{Beane11a}, yields the dashed curve. 
It is obvious that the dependence on $M_\pi$ we obtain agrees -- at least 
on a qualitative level -- with that presented in Ref.~\cite{Beane11}. 
Specifically, our calculation exhibits the same trend (a decrease of the binding 
energy with decreasing pion mass) and our binding energy of 9 MeV at the physical 
pion mass is within the error bars of the results given in~\cite{Beane11}. 
On the other hand, we clearly observe a non-linear dependence of the binding
energy on the pion mass. As a consequence, scaling our results to the binding
energy reported by the HAL QCD Collaboration \cite{Inoue} (30-40 MeV for  
$M_\pi \approx 700-1000$~MeV) yields binding energies of more than 20 MeV
at the physical point, which is certainly outside of the range suggested in
Ref.~\cite{Beane11}. However, we note that for such large pion masses the LO
chiral EFT can not be trusted quantitatively. 
We remark that in our simulations the curves corresponding to different binding 
energies remained roughly parallel even up to such large values as suggested 
by the HAL QCD Collaboration. 

Our finding that any $H$-dibaryon is very likely a bound $\Xi N$ state rather than a $\La\La$
state, which follows from the assumed (approximate) SU(3) symmetry of the interaction, 
suggests that not only the pion mass but also the masses of the baryons play a significant 
role for the concrete value of binding energy. 
In the specific case of $J=0$, $I=0$, $S=-2$ we are dealing with three coupled channels, 
namely $\La\La$, $\Xi N$, and $\Si\Si$. For the isospin-averaged masses that we use 
their thresholds are at 2231.2, 2257.7, and 2385.0~MeV, respectively. 
Thus, the physical difference between the $\La\La$ and $\Xi N$ thresholds of around
26~MeV implies that the $H$-dibaryon considered above is, in reality, bound by roughly
28~MeV with respect to its ``proper'' threshold. Accordingly, one intuitively expects that 
in a fully SU(3) symmetric case, where the masses of all octet baryons coincide, the 
bound state would remain more or less fixed to the $\Xi N$ threshold and then would lie also 
about 28 MeV below the $\La\La$ threshold. 
Since we know from our experience with coupled-channel problems 
\cite{Polinder06,Hai10a,Hai05,Hai11} that coupling effects are sizeable and 
the actual separation of the various thresholds plays a crucial role we investigated also
the dependence of the $H$ binding energy on the thresholds (i.e. on the $\Sigma$, and on 
the $\Xi$ and $N$ masses). Corresponding results are displayed in Fig.~\ref{fig:mpi} in
the right panel. 

We start with considering the effect of the $\Sigma \Sigma$ channel because its threshold 
is separated by roughly 154 MeV from the one of $\La\La$ so that there is a rather
drastic breaking of the SU(3) symmetry. Indeed,
when we decrease the $\Sigma$ mass so that the nominal $\Sigma \Sigma$ threshold 
(at 2385~MeV) moves downwards and finally coincides with the one of the 
$\La\La$ channel (2231.2~MeV), we observe a concurrent fairly drastic 
increase in the $H$ binding energy, cf. the solid curve in Fig.~\ref{fig:mpi} for
results based on the interaction YY-D with a binding energy of -1.87~MeV for 
physical masses of the mesons and baryons. 
In this context we want to point out that the direct
interaction in the $\Sigma\Sigma$ channel is actually repulsive for the low-energy
coefficients $C^{27}$ and $C^{8_s}$ fixed from the $YN$ data plus the pseudoscalar
meson exchange contributions with coupling constants determined from the SU(3)
relations Eq.~(\ref{su3}), and it remains repulsive even for $C^{1}$ values that produce
a bound $H$-dibaryon. But the coupling between the channels generates a sizeable
effective attraction which increases when the channel thresholds come closer. 
The dashed curve is a calculation with the contact term $C^{1}$ fixed to simulate 
the binding energy ($-13.2$~MeV) of the NPLQCD Collaboration at $M_\pi=389\,$MeV.
As one can see, the dependence of the binding energy on the $\Sigma$ mass is rather
similar. The curve is simply shifted downwards by around 4.5~MeV, i.e. by the difference
in the binding energy observed already at the physical masses. 
The asterisks and crosses represent results where, besides the variation of the 
$\Sigma\Sigma$ threshold, the $\Xi N$ threshold is shifted to coincide with that
of the $\La\La$ channel. This produces an additional increase of the $H$ binding 
energy by 20~MeV at the physical $\Sigma\Sigma$ threshold and by 9~MeV 
for that case where all three $BB$ threshold coincide. 
Altogether there is an increase in the binding energy of roughly 60~MeV 
when going from the physical point to the case of baryons with identical
masses. This is significantly larger than the variations due to the pion mass
considered before. Note that we have kept the pion mass at its physical value
while varying the $BB$ thresholds. 

\subsection{Comparison with lattice QCD results}
 
After these exemplary studies let us now try to connect with the published 
$H$ binding energies from the lattice QCD calculations \cite{Beane11a,Inoue}. 
The results obtained by the HAL QCD Collaboration are obviously for the SU(3)
symmetric case and the corresponding masses are given in Table I of 
Ref.~\cite{Inoue}. Thus, we can take those masses and then fix the LEC
$C^1$ so that we reproduce their $H$ binding energy with those masses. 
To be concrete: we use $M_{ps} = 673\,$MeV and $m_{B} = 1485\,$MeV, and fix 
$C^1$ so that $E_H = -35\,$MeV. We denote this interaction by YY-HAL.
When we now let the masses of the baryons and
mesons go to their physical values the bound state moves up to the
$\La\La$ threshold, crosses the threshold, crosses also the $\Xi N$ threshold and 
then disappears. In fact, qualitatively this outcome
can be already read off from the curves in Fig.~\ref{fig:mpi} by combining the
effects from the variations in the pion and the baryon masses. Based on those
results one expects a shift of the $H$ binding energy in the order of
60 to 70~MeV for the mass parameters of the HAL QCD calculation. 

In case of the NPLQCD calculation we take the values provided in 
Ref.~\cite{Beane11b}, as before. Those yield
then 17~MeV for the $\Xi N$-$\La\La$ threshold separation 
(to be compared with the physical value of roughly 26~MeV) and 
77~MeV for the $\Si\Si$-$\La\La$ separation (physical value around 154~MeV). 
We also use the meson masses of Ref.~\cite{Beane11b}, specifically
$M_\pi = 389$ MeV. 
With those baryon and meson masses we fix again the LEC $C^1$ so that we 
reproduce the $H$ binding energy given by the NPLQCD Collaboration, namely 
$E_H =-13.2$~MeV \cite{Beane11a} (called YY-NPL in the following). 
Again we let the masses of the baryons and mesons approach their 
physical values. Also here the bound state moves up to and crosses
the $\La\La$ threshold. However, in the NPLQCD case the state survives
and remains below the $\Xi N$ threshold at the physical point. Specifically,
we observe a resonance at a kinetic energy of 21~MeV in the $\La\La$
system or, more precisely, a quasi-bound state in the $\Xi N$ system around 
5~MeV below its threshold. 

\begin{figure}[t!]
\centering
\includegraphics[width=0.305\textwidth,keepaspectratio,angle=-90]{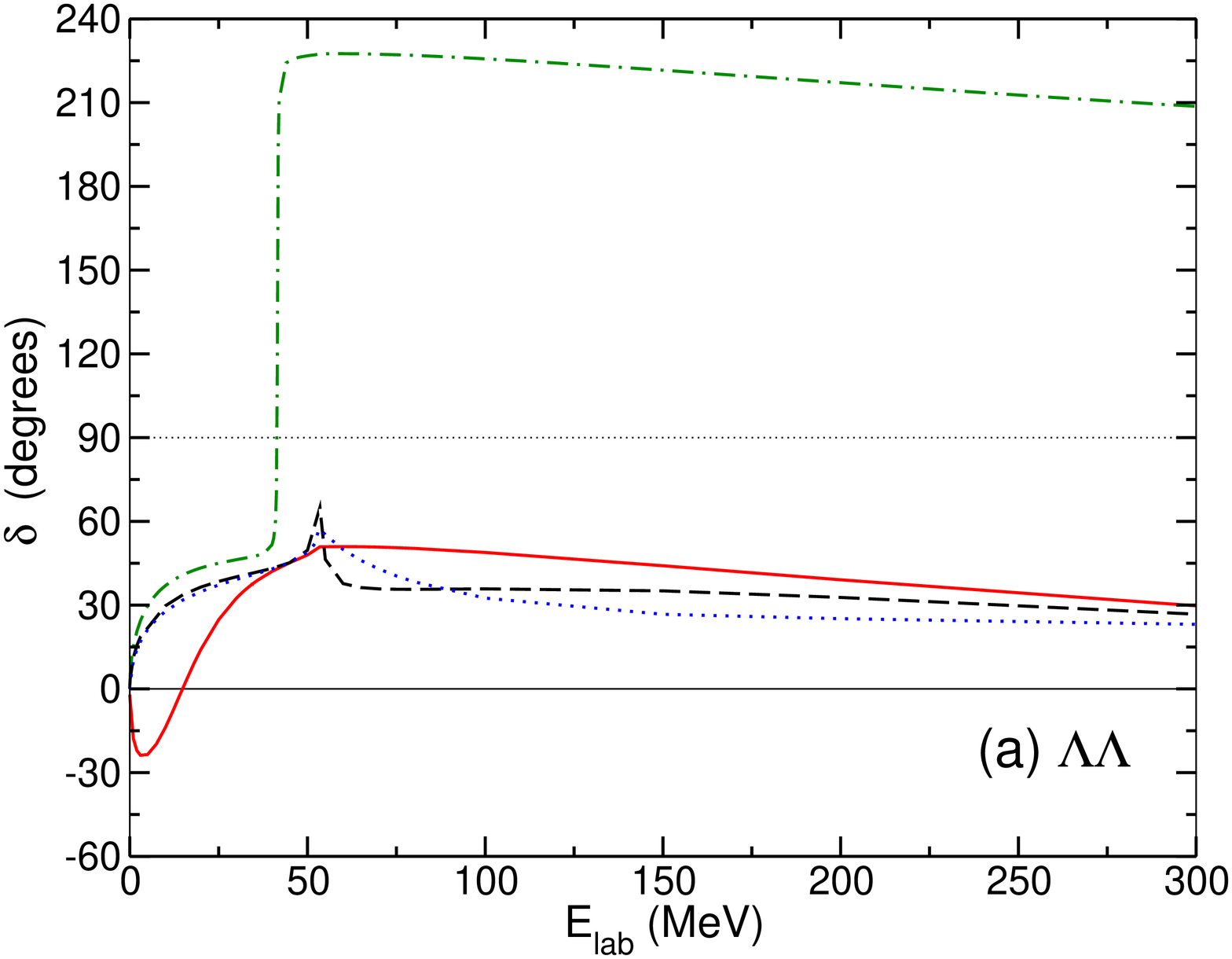}
\includegraphics[width=0.305\textwidth,keepaspectratio,angle=-90]{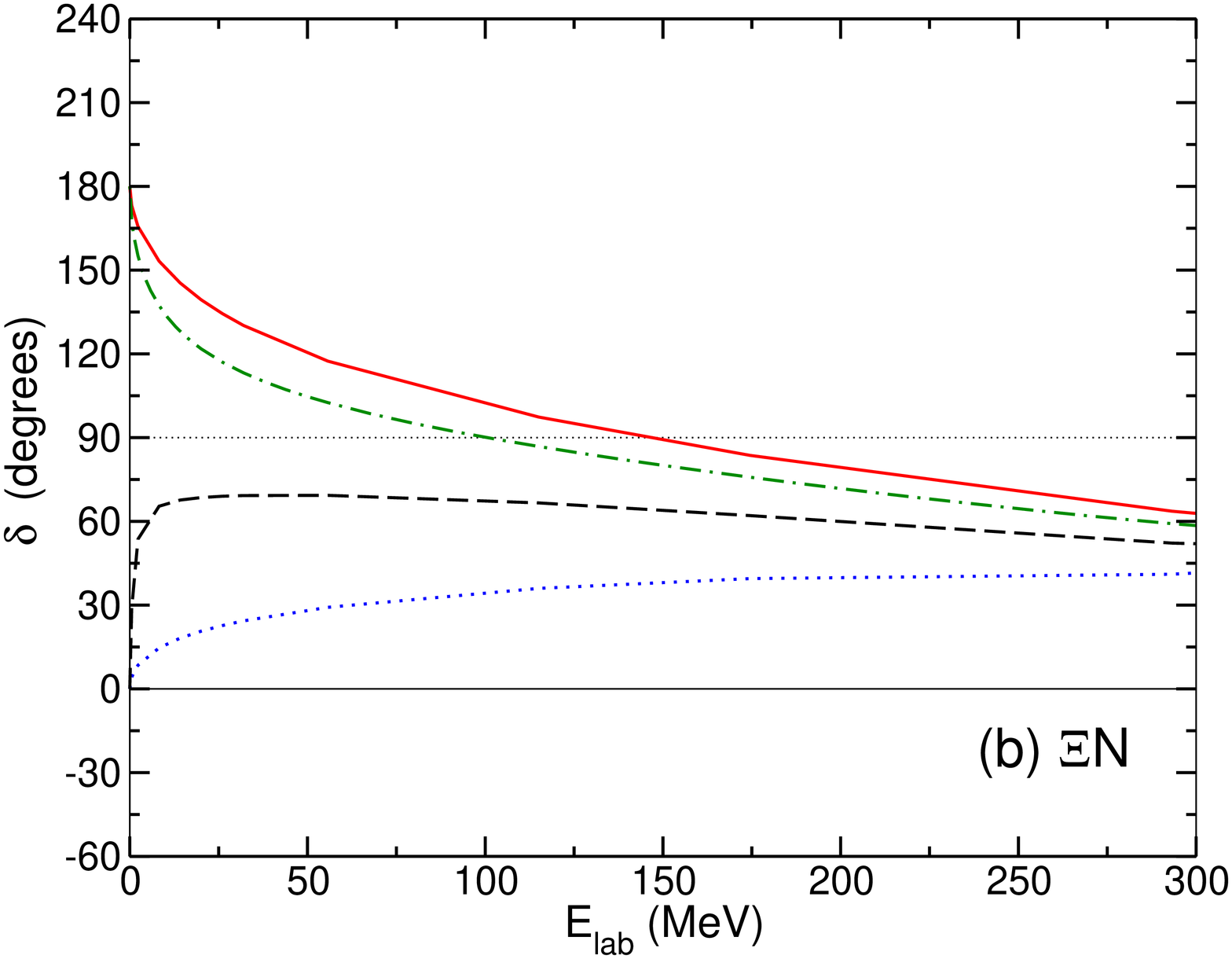}
\includegraphics[width=0.305\textwidth,keepaspectratio,angle=-90]{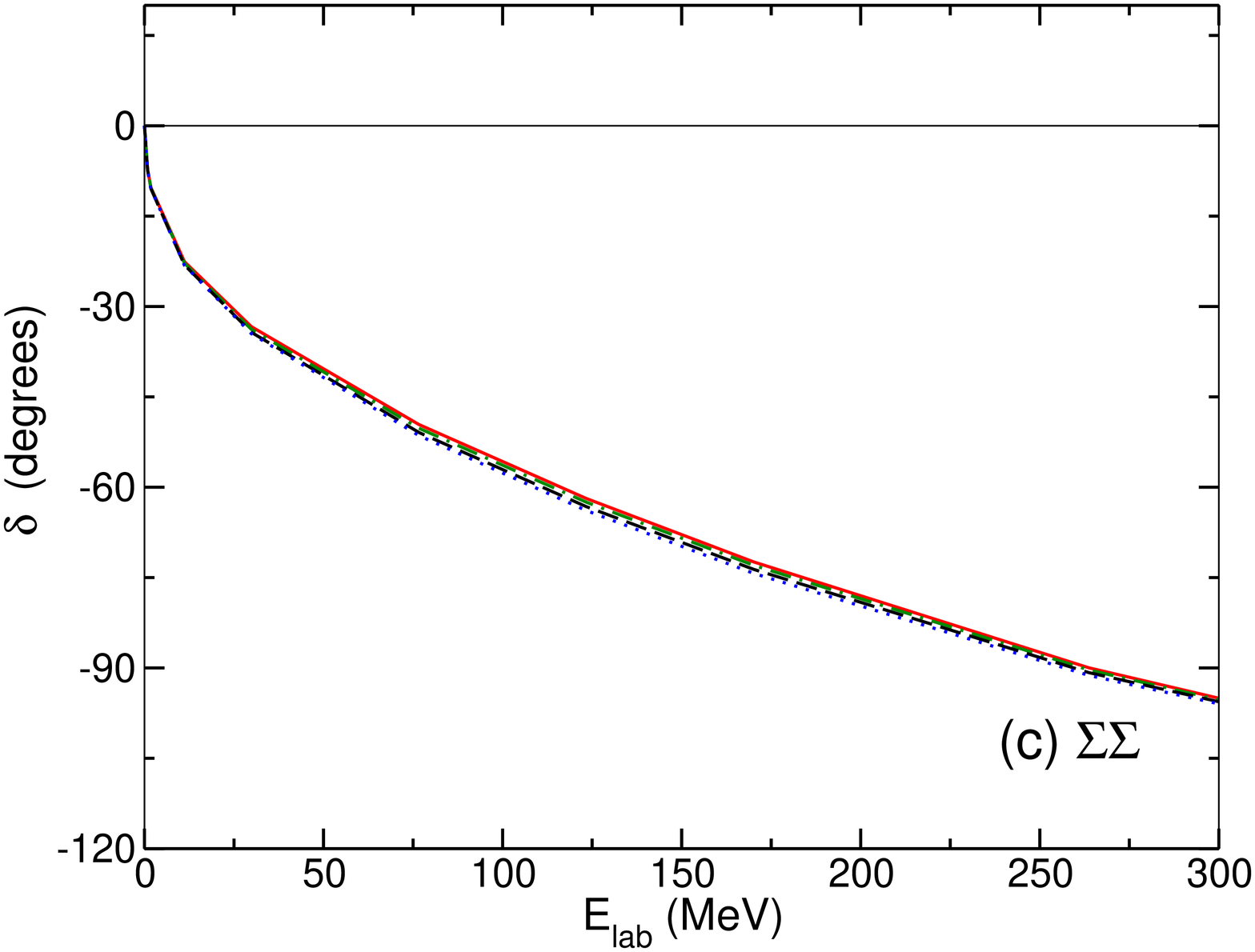}
\caption{Phase shifts in the $^1S_0$ partial wave in the $I=0$ channel of $\La\La$ (a), 
$\Xi N$ (b) and $\Si\Si$ (c) as a function of the pertinent laboratory energies. 
The solid line is the result for our illustrative $BB$ interaction that produces a bound $H$ at
$E_H=-1.87\,$MeV. 
The dotted line corresponds to the EFT potential of Ref.~\cite{Polinder07}
(Table 4) with cutoff mass $\Lambda = 550$ MeV.
The other curves are results for interactions that are fine-tuned to the 
$H$ binding energies found in the lattice QCD calculations of the HAL QCD (dashed) and 
NPLQCD (dash-dotted) Collaborations, respectively, for the pertinent meson (pion) and
baryon masses as described in the text. 
}
\label{fig:phases}
\end{figure}

It is interesting to observe that the chiral extrapolation of the lattice
QCD results performed by Shanahan et al. \cite{Shanahan11} yields results
that are qualitatively similar to ours. In that reference the authors conclude 
that the $H$-dibaryon is likely to be unbound by 13$\pm$14 MeV at the physical point. 
Let us emphasize, however, that our values are not really comparable
with theirs. As said above, in our analysis we assume that the $H$-dibaryon is actually 
a bound $BB$ state -- which seems to be the case also in the lattice QCD
studies \cite{Beane11a,Inoue}. On the other hand, in Ref.~\cite{Shanahan11} it is 
assumed that the $H$ is a compact, multi-quark state rather than a loosely bound
molecular state, i.e. an object as originally suggested by Jaffe. 
How such a genuine multi-quark state would be influenced
by variations of the $BB$ thresholds is completely unclear. It depends,
among other things, on whether and how strongly this state couples to the
$\La\La$, $\Xi N$, and $\Si\Si$ channels. So far there is no information
on this issue from lattice QCD calculations. Clearly, in case of a strong and 
predominant coupling to the $\La\La$ alone, variations of the $\Si\Si$ and
$\Xi N$ would not influence the $H$ binding energy significantly. However,
should it couple primarily to the $\Xi N$ and/or $\Si\Si$ channels
then we expect a sensitivity of the binding energy to their thresholds values
comparable to what we found in our study for the case of a bound state. 

Phase shifts for the $^1S_0$ partial wave of the $\La\La$, $\Xi N$ and 
$\Sigma\Sigma$ channels are presented in Fig.~\ref{fig:phases}. 
The solid line is the result for the $BB$ 
interaction YY-D that produces a loosely bound $H$ dibaryon with $E_H=-1.87$~MeV.
The phase shift for the $\Xi N$ channel, Fig.~\ref{fig:phases} (b), is rather similar to the 
one for the $^3S_1$ $NN$ partial wave where the deuteron resides, 
see e.g. \cite{Epe05}. 
Specifically, it starts at $180^o$, decreases smoothly and eventually 
approaches zero for large energies, fulfilling the Levinson theorem. 
The result for $\La\La$ ($^1S_0$), Fig.~\ref{fig:phases} (a), behaves rather differently. 
This phase commences at zero degrees, is first negative but becomes positive within 
20 MeV and finally turns to zero again for large energies.
The dashed curve corresponds to the interaction YY-HAL that was fitted to the 
result of the HAL QCD Collaboration and reproduces their bound $H$ dibaryon with
their meson and baryon masses. 
The phase shift of the $\Xi N$ channel, calculated with physical masses, 
shows no trace of a bound state anymore. Still the phase shift 
rises up to around $60^o$ near threshold, a behavior quite similar to that of 
the $^1S_0$ $NN$ partial wave where there is a virtual state (also called
antibound state \cite{Kok,Pearce}). Indeed, such a virtual state also seems to 
be present in the $\Xi N$ channel as a remnant of the original bound state.
The effect of this virtual state can be seen in the $\La\La$ phase shift where it
leads to an impressive cusp at the opening of the $\Xi N$ channel, 
cf. the dashed line in Fig.~\ref{fig:phases} (a). 

In the $\Xi N$ phase shifts for the NPLQCD case (dash-dotted curve) 
the presence of a bound state is clearly visible. The corresponding $\La\La$
phase shift exhibits a resonance-like behavior at the energy where the
(quasi-bound) $H$ dibaryon is located.  

The dotted curves are the results for the original chiral EFT potential
with cut-off $\Lambda = 550$ MeV as published in \cite{Polinder07}. The
$\La\La$ as well as the $\Xi N$ phase shifts are qualitatively similar
to the ones for the HAL QCD case. But the smaller $\Xi N$ phase shift
together with the reduced cusp effect indicate that there is no 
near-by virtual state produced by this interaction. 

The $\Sigma\Sigma$ phase shifts predicted by the various interactions are
almost the same, cf. Fig.~\ref{fig:phases} (c). This may be not too surprising. 
After all, the 
$\Sigma\Sigma$ threshold is rather far away from the one of the 
$\Lambda\Lambda$ channel and the region, where we have introduced the $H$ 
dibaryon. Thus, it remains practically unaffected by those changes. 

\begin{figure}[t!]
\centering
\includegraphics[width=0.305\textwidth,keepaspectratio,angle=-90]{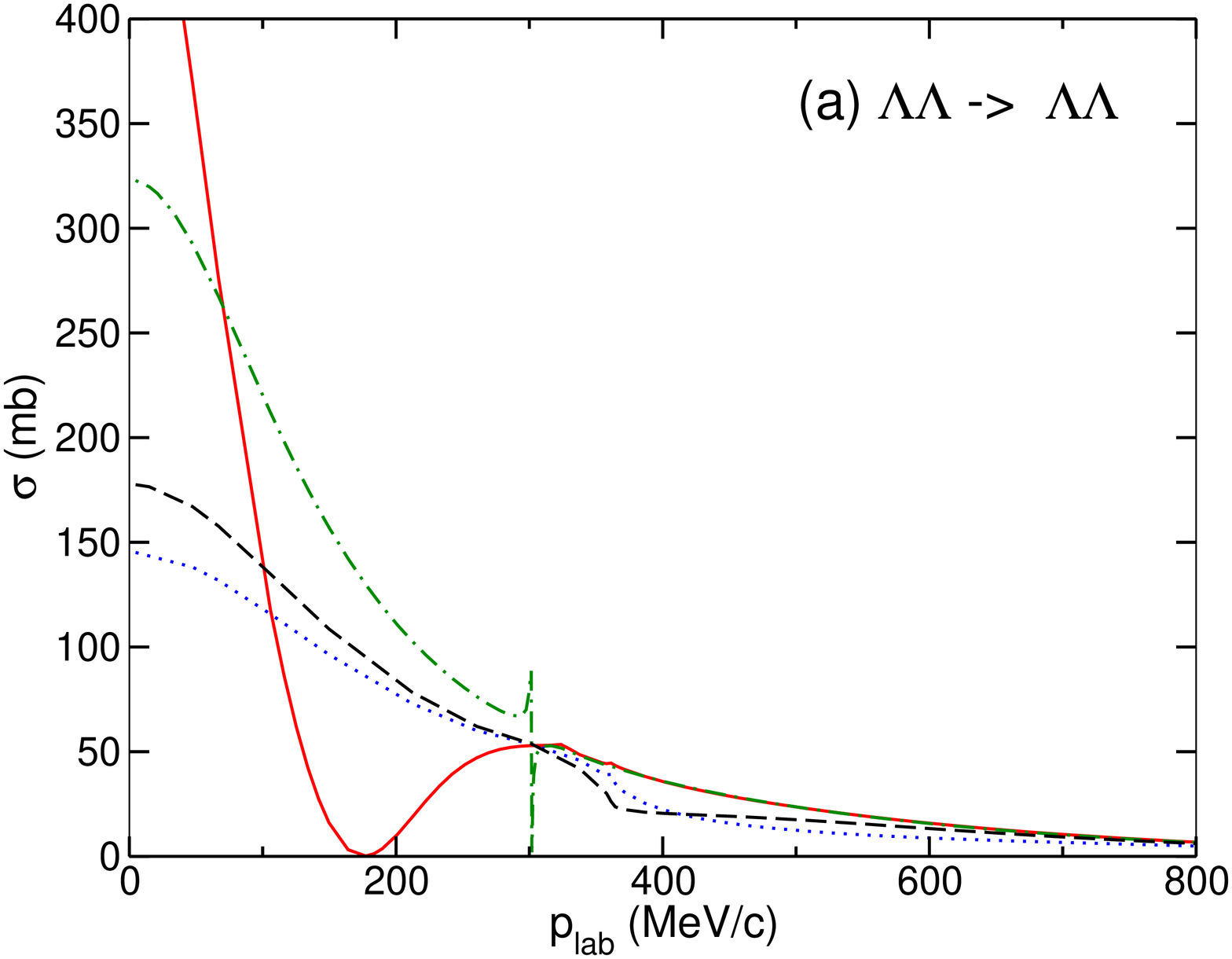}
\includegraphics[width=0.305\textwidth,keepaspectratio,angle=-90]{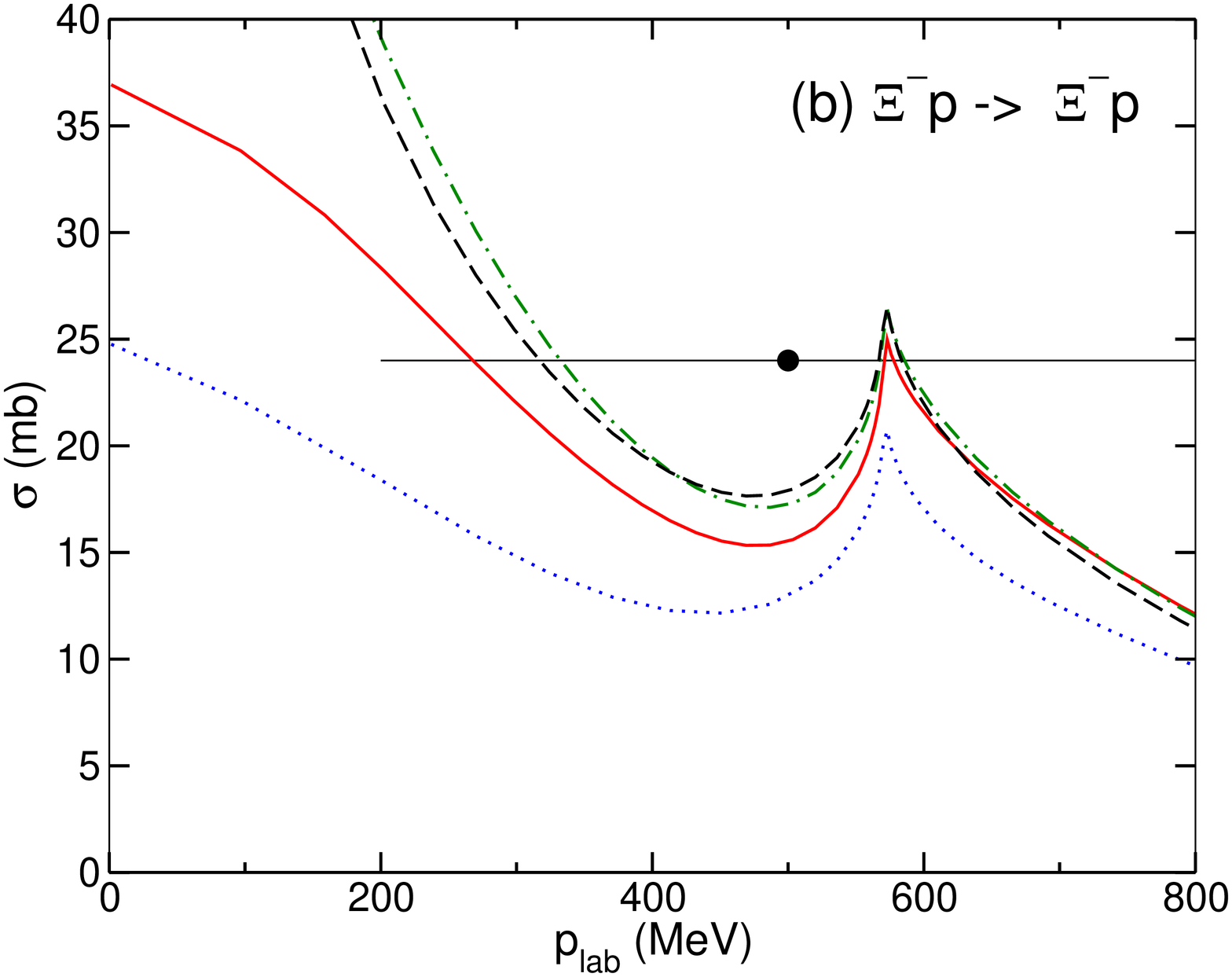}
\includegraphics[width=0.305\textwidth,keepaspectratio,angle=-90]{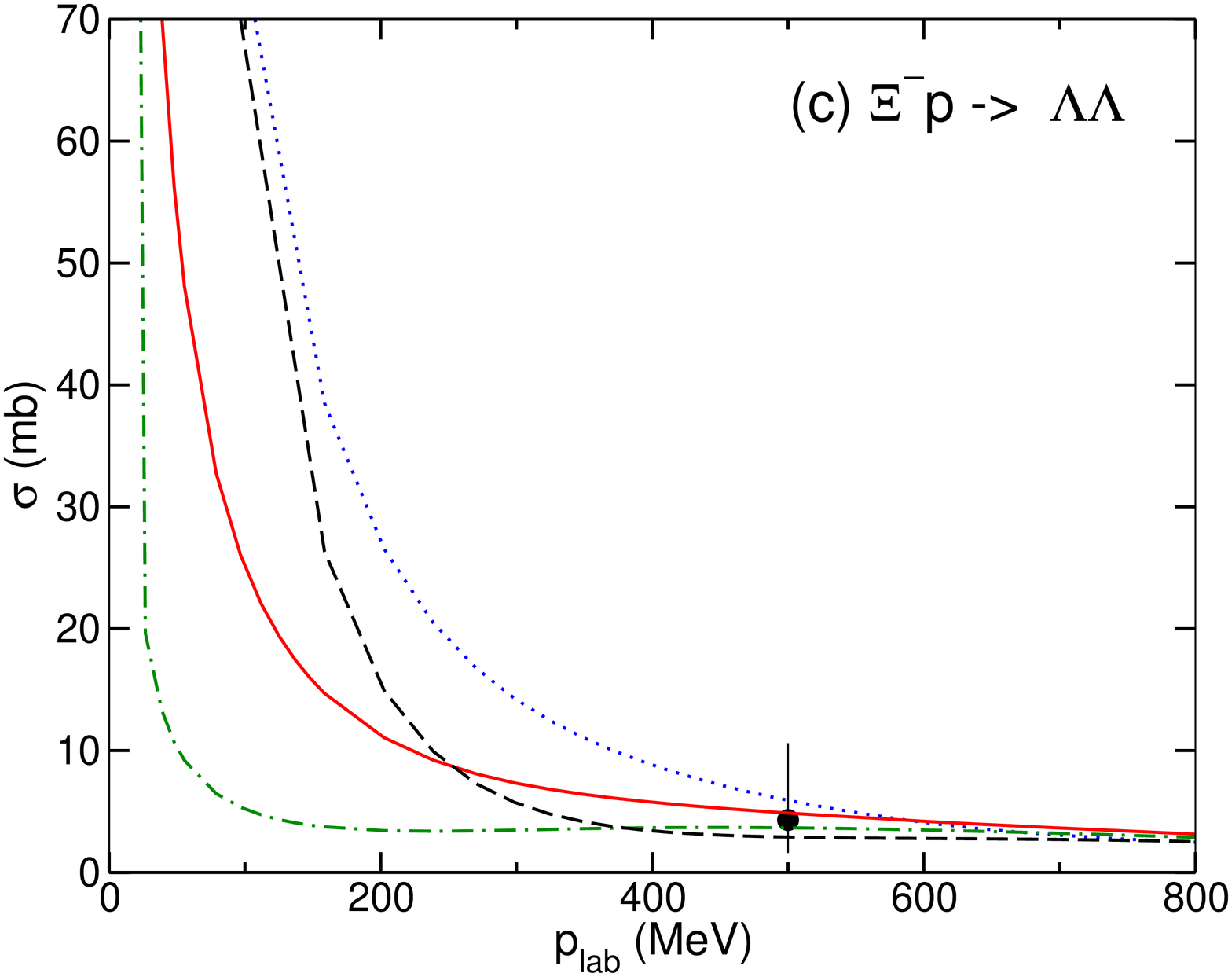}
\caption{Total cross sections for some $S=-2$ channels as a function of $p_{lab}$. 
The solid line is the result for our illustrative $BB$ interaction that produces a bound $H$ at
$E_H=-1.87$ MeV. The dotted line corresponds to the EFT potential of Ref.~\cite{Polinder07}
(Table 4) with cutoff mass $\Lambda = 550$ MeV.
The other curves are results for interactions that are fine-tuned to the 
$H$ binding energies found in the lattice QCD calculations of the HAL QCD (dashed) and 
NPLQCD (dash-dotted) Collaborations, respectively, for the pertinent meson (pion) and
baryon masses as described in the text. 
The experimental cross sections in (b) and (c) are taken from Ref. \cite{Ahn:2005jz}.
}
\label{fig:cross}
\end{figure}

Finally, for illustrative purposes, we present cross sections for the 
$\La\La$ and $\Xi^- p$ channels. Results of corresponding calculations, 
now performed in particle basis (but neglecting the Coulomb interaction), 
are displayed in Fig.~\ref{fig:cross}. There are some experimental constraints 
for these two channels. In particular, there is an upper limit of $24$ mb at $90\%$ 
confidence level for elastic $\Xi^-p$ scattering, 
while for the $\Xi^-p\rightarrow \Lambda\Lambda$ cross section at $p_{\rm lab}=500$ MeV/c 
a value of $4.3^{+6.3}_{-2.7}$ mb was reported \cite{Ahn:2005jz}.

As obvious from Fig.~\ref{fig:cross}, there are significant differences in the
cross sections predicted by the $YY$ interactions generated in the context of 
the $H$-dibaryon discussion -- however, only at low momenta where no
experimental information is at hand so far. Anyhow, those results suggest that
a determination of the $\Xi^- p$ cross section at $p_{\rm lab} \approx 200$ MeV/c,
say, with reasonable errors would already put strong constraints on the
$H$-dibaryon. In particular, situations where it is located close to the $\Xi N$
threshold -- as at is the case in our simulations of the NPLQCD and HAL QCD results 
(dash-dotted and dashed curves, respectively) -- could be ruled out. 
Distinguishing an actually (though loosely) bound $H$-dibaryon (solid curve) from 
the situation without any $H$-dibaryon (dotted curve) certainly requires better
statistics. Here one has to keep in mind that the $\Xi^-p$ cross section would 
be even closer to the latter result, should the $H$-dibaryon be somewhat 
stronger bound than assumed in our calculation. Note that the peak in the
$\Xi^- p$ cross section around 575 MeV/c is a cusp due to the opening of the
$\Sigma^0\Lambda$ channel. 

There are also characteristic differences in the predictions for
$\Xi^- p\rightarrow \Lambda\Lambda$, cf. Fig.~\ref{fig:cross} (c). 
However, since this cross section rises to infinity with decreasing $p_{\rm lab}$,
due to the phase-space factor, it might be more difficult to draw conclusions
in this case. 

The assumed $H$ dibaryon below the $\La\La$ threshold introduces a rather 
strong and peculiar energy variation in the near-threshold $\La\La$ cross 
section, cf. the solid curve in Fig.~\ref{fig:cross} (a). 
The effects due to the other considered interactions 
is less spectacular, specifically, because the structure produced by the NPLQCD
case is so narrow that it would be presumably completely washed out once one
takes into account the finite momentum resolution of an actual experiment. 
In any case, measuring the $\La\La$ cross section directly seems to be 
practically impossible. However, one could measure the $\La\La$ invariant mass
spectrum in reactions like $K^- A \to \La\La + X$ where $A$ can be the deuteron 
or a heavier nucleus. As a matter of facts, corresponding results from a measurement 
of $K^-\phantom{}^{12}\rm C$ have been already published \cite{Yoon}. 
Still, it is unclear whether such an invariant mass distribution would
be dominated by the $\La\La \to \La\La$ transition amplitude or rather by 
$\Xi N \to \La\La$. Since our investigation suggests that any near-threshold
$H$-dibaryon will have a large if not dominant $\Xi N$ component
one expects that then the $\Xi N \to \La\La$ amplitude should play 
likewise an important if not decisive role for the $\La\La$ 
invariant mass distribution. 

\section{Summary}

In this paper we have presented an analysis of the quark mass dependence of 
binding energies for baryon-baryon systems in the strangeness $S=-2$, 
$S=-3$, and $S=-4$ sectors in the framework of chiral effective field 
theory at leading order in the Weinberg counting. 
In particular, we have explored the dependence of those binding energies 
on the pion mass in order to connect with current lattice QCD calculations.
We remark that at higher orders, other effects like the quark mass dependence of 
the meson-baryon couplings or of the contact interactions will have to be
considered (see e.g.~\cite{Epe02}).

With regard to the $\Xi\Xi$, $\Xi\Sigma$ and $\Xi\Lambda$ systems, where
meson-exchange potentials as well as leading-order EFT interactions, 
derived under the assumption of (broken) SU(3) symmetry, predict
the existence of bound states in the various $^1S_0$ partial waves, we 
find a rather weak dependence of the binding energies on the pion mass.
For the $\Xi^-\Xi^-$ system a calculation performed with meson and
baryon masses that match the status of a recent lattice QCD exploration
by the NPLQCD Collaboration yields binding energies that are 
compatible with the reported lattice QCD result \cite{Beane11a} 
within the given error bars. 
 
We have also investigated the situation concerning the so-called $H$-dibaryon.  
Here we found rather drastic effects caused by the SU(3) breaking related 
to the values of the three thresholds $\Lambda\Lambda$, $\Sigma\Sigma$ and $\Xi N$.
For physical values the binding energy of the $H$ is reduced by as much as 60~MeV 
as compared to a calculation based on degenerate (i.e. SU(3) symmetric) $BB$ 
thresholds. 
Translating this observation to the situation in the HAL QCD \cite{Inoue} 
calculation, we see that the bound state has disappeared at the physical point. 
For the case of the NPLQCD calculation \cite{Beane11a}, 
a resonance in the $\Lambda\Lambda$ system might survive.

\ack
J.H. acknowledges stimulating discussions with N.N. Nikolaev.
This work is supported by the EU-Research
Infrastructure Integrating Activity ``Study of Strongly Interacting Matter''
(HadronPhysics2, grant n. 227431) under the Seventh Framework Program of the EU,
and by the DFG (SFB/TR 16 ``Subnuclear Structure of Matter'').


\bigskip 

\begin{thebibliography}{10}
\bibitem{Jaffe:1976yi}
R. L. Jaffe, Phys. Rev. Lett. {\bf 38}, 195 (1977) [Erratum-ibid. {\bf 38}, 617 (1977)]. 
\bibitem{Adlarson}
  P.~Adlarson {\it et al.} [ WASA-at-COSY Collaboration ],
  Phys.\ Rev.\ Lett.\  {\bf 106}, 242302 (2011).
\bibitem{Gal10}
  A.~Gal, 
  [arXiv:1011.6322 [nucl-th]].
\bibitem{Miller}
  G. Miller, 
  [arXiv:nucl-th/0607006].

\bibitem{Stoks:1999bz}
  V.~G.~J.~Stoks and T.~A.~Rijken,
  Phys.\ Rev.\  C {\bf 59}, 3009 (1999).

\bibitem{Polinder06}
  H.~Polinder, J.~Haidenbauer, U.-G.~Mei{\ss}ner,
  Nucl.\ Phys.\  A {\bf 779}, 244 (2006).
\bibitem{Polinder07}
  H.~Polinder, J.~Haidenbauer, U.-G.~Mei{\ss}ner,
  Phys.\ Lett.\ B {\bf 653}, 29 (2007).
\bibitem{Hai10a}
  J.~Haidenbauer, U.-G.~Mei{\ss}ner,
  Phys.\ Lett.\ {\bf B684}, 275 (2010).

\bibitem{Yoon}
C.J. Yoon {\it et al.}, Phys. Rev. C {\bf 75}, 022201 (2007). 

\bibitem{Beane}
  S.~R.~Beane {\it et al.},
  Phys.\ Rev.\ Lett.\  {\bf 106}, 162001 (2011).
\bibitem{Beane11a}
 S.~R.~Beane {\it et al.},
  [arXiv:1109.2889 [hep-lat]].
\bibitem{Inoue}
  T.~Inoue {\it et al.},, 
  Phys.\ Rev.\ Lett.\  {\bf 106}, 162002 (2011).
\bibitem{Inoue11a}
  T.~Inoue, 
  [arXiv:1109.1620 [hep-lat]].

\bibitem{Beane11}
 S.~R.~Beane {\it et al.},
  [arXiv:1103.2821 [hep-lat]].
\bibitem{Shanahan11}
   P.~E.~Shanahan, A.~W.~Thomas, R.~D.~Young, 
   Phys.\ Rev.\ Lett.\  {\bf 107}, 092004 (2011).

\bibitem{Beane:2002vs}
  S.~R.~Beane, M.~J.~Savage,
  Nucl.\ Phys.\ A {\bf 713}, 148 (2003).
\bibitem{Beane03}
  S. R. Beane, M. J. Savage, Nucl. Phys. A {\bf 717}, 91 (2003).
\bibitem{Epe02}
  E.~Epelbaum, U.-G.~Mei{\ss}ner, W.~Gl\"ockle,
  Nucl.\ Phys.\ A {\bf 714}, 535 (2003).
\bibitem{Epe02a}
E.~Epelbaum, U.-G.~Mei{\ss}ner, W.~Gl\"ockle, [arXiv:nucl-th/0208040]. 

\bibitem{Ahn:2005jz}
  J.~K.~Ahn {\it et al.},
  Phys.\ Lett.\  B {\bf 633}, 214 (2006).

\bibitem{Haidenbauer11}
  J.~Haidenbauer, U.~-G.~Mei{\ss}ner,
  Phys.\ Lett.\ B {\bf 706}, 100 (2011).

\bibitem{Haidenbauer07}
  J.~Haidenbauer, U.~-G.~Mei{\ss}ner, A.~Nogga, H.~Polinder,
  Lect.\ Notes Phys.\  {\bf 724}, 113 (2007).

\bibitem{Savage1}
  M.~J.~Savage, M.~B.~Wise,
  Phys.\ Rev.\ D {\bf 53}, 349 (1996).

\bibitem{Korpa}
  C.~L.~Korpa, A.~E.~L.~Dieperink, R.~G.~E.~Timmermans,
  Phys.\ Rev.\ C {\bf 65}, 015208 (2002).

\bibitem{Savage2}
  S.~R.~Beane, P.~F.~Bedaque, A.~Parre\~no, M.~J.~Savage,
  Nucl.\ Phys.\ A {\bf 747}, 55 (2005). 

\bibitem{Swart}
J.~J.~de Swart, Rev. Mod. Phys. {\bf 35}, 916 (1963). 
\bibitem{Dover}
C.~B.~Dover, H.~Feshbach, Annals Phys. {\bf 217}, 51 (1992). 

\bibitem{Ratcliffe} P.~G.~Ratcliffe, Phys. Lett. B {\bf  365}, 383 (1996).

\bibitem{Yamanishi} T.~Yamanishi, Phys. Rev. D {\bf 76}, 014006 (2007).

\bibitem{Epe98}
  E.~Epelbaum, W.~Gl\"ockle, U.-G.~Mei{\ss}ner,
  Nucl.\ Phys.\ A {\bf 637}, 107 (1998).
\bibitem{Epe05}
  E.~Epelbaum, W.~Gl\"ockle, U.-G.~Mei{\ss}ner,
  Nucl.\ Phys.\ A {\bf 747}, 362 (2005).

\bibitem{Takahashi:2001nm}
  H.~Takahashi {\it et al.},
  Phys.\ Rev.\ Lett.\  {\bf 87}, 212502 (2001).

\bibitem{Gal}
  I.~N.~Filikhin, A.~Gal, V.~M.~Suslov, 
  Phys.\ Rev.\ C {\bf 68}, 024002 (2003).

\bibitem{Rijken}
  T.~A.~Rijken, Y.~Yamamoto,
  Phys.\ Rev.\ C {\bf 73}, 044008 (2006).

\bibitem{Fujiwara}
  Y.~Fujiwara, Y.~Suzuki, C.~Nakamoto,
  Prog.\ Part.\ Nucl.\ Phys.\  {\bf 58}, 439 (2007).

\bibitem{Ashot}
  A.M. Gasparyan, J. Haidenbauer, C. Hanhart, 
  [arXiv:1111.0513 [nucl-th]].

\bibitem{Frink:2005ru}
  M.~Frink, U.-G.~Mei{\ss}ner, I.~Scheller,
  Eur.\ Phys.\ J.\  {\bf A24 }, 395 (2005).

\bibitem{Beane11b}
  S.~R.~Beane et al., 
  Phys.\ Rev.\ D {\bf 84}, 014507 (2011).

\bibitem{Schaffner1999}
  J.~Schaffner-Bielich, R.~Mattiello and H.~Sorge,
  Phys.\ Rev.\ Lett.\  {\bf 84}, 4305 (2000).

\bibitem{Steinheimer}
  J.~Steinheimer, M.~Mitrovski, T.~Schuster, H.~Petersen, M.~Bleicher, 
  H.~St\"ocker, 
  Phys.\ Lett.\  B {\bf 676}, 126 (2009).

\bibitem{Schwinger47}
J.~Schwinger,
Phys.\ Rev.\ {\bf 72}, 742 (1947).

\bibitem{Bethe}
  H.~A.~Bethe, Phys.\ Rev.\ {\bf 76}, 38 (1949).

\bibitem{Oka}
  M. Oka, K. Shimizu, K. Yazaki,
  Phys.\ Lett.\ {\bf 130B}, 365 (1983). 

\bibitem{Nakamoto}
  C.~Nakamoto, Y. Suzuki, Y. Fujiwara, 
  Prog. Theor. Phys. {\bf 97}, 761 (1997). 

\bibitem{Hai05}
  J.~Haidenbauer, U.-G.~Mei{\ss}ner,
  Phys.\ Rev.\ C {\bf 72}, 044005 (2005).
\bibitem{Hai11}
  J.~Haidenbauer, G.~Krein, U.-G.~Mei{\ss}ner, L.~Tolos,
  Eur.\ Phys.\ J.\ A {\bf 47}, 18 (2011).

\bibitem{Kok} A.~M.~Badalyan, L.~P.~Kok, M.~I.~Polikarpov, Yu.~A.~Simonov, 
  Phys. Rept. {\bf 82}, 31 (1982). 
\bibitem{Pearce} 
  B.~C.~Pearce, B.~F.~Gibson,
  Phys.\ Rev.\  C {\bf 40}, 902 (1989).

\end{thebibliography}
\end{document}